\documentclass[12pt]{article}
\usepackage{hyperref}
\hypersetup{
colorlinks=true,
linkcolor=black,
citecolor=black,
urlcolor=blue,
filecolor=black,
pdfborder={0 0 0},
}

\usepackage{color}
\definecolor{darkblue}{rgb}{0,0,0.5}
\definecolor{darkred}{rgb}{0.5,0,0}
\definecolor{darkgreen}{rgb}{0,0.5,0}
\usepackage{hyperref}
\hypersetup{
colorlinks=true,
linkcolor=black,
citecolor=black,
urlcolor=darkblue,
filecolor=black,
pdfborder={0 0 0},
}
\usepackage{amssymb}
\usepackage{amsfonts}
\usepackage{accsupp}
\usepackage{mathrsfs}
\usepackage{latexsym}
\usepackage{amsmath}
\usepackage{graphics}
\usepackage{graphicx}
\usepackage{sectsty}
\usepackage{setspace}
\usepackage{multirow}
\sectionfont{\normalsize\bf}
\subsectionfont{\rm\normalsize}
\def\be{\begin{equation}}
\def\ee{\end{equation}}
\input amssym.def
\input amssym
\def\cL{{\cal L}}

\def\T{{\cal T}}

\newcount\nx
\newcount\nxo
\newcount\ny
\newcount\nyo
\newcount\nxd
\newcount\nxod
\newcount\nyd
\newcount\nyod
\newcount\nl

\def\grid#1#2#3#4{
\nxo=#1
\nyo=#2
\nx=#3
\ny=#4    
\multiply \nxo by 4
\multiply \nyo by 4
\advance \nx by 1
\multiply \ny by 4
\linethickness{0.075mm}
\multiput(\nxo,\nyo)(4,0){\nx}{\line(0,1){\ny}}
\nx=#3
\ny=#4
\advance \ny by 1
\multiply \nx by  4
\multiput(\nxo,\nyo)(0,4){\ny}{\line(1,0){\nx}}
}

\def\dgrid#1#2#3#4{
\nxo=#1
\nyo=#2
\nx=#3
\ny=#4
\multiply \nxo by 4
\multiply \nyo by 4
\advance \nxo by - 2
\advance \nyo by - 2
\advance \nx by 1
\multiply \ny by  4
\linethickness{0.3mm}
\multiput(\nxo,\nyo)(4,0){\nx}{\line(0,1){\ny}}
\nx=#3
\ny=#4
\advance \ny by 1
\multiply \nx by  4
\multiput(\nxo,\nyo)(0,4){\ny}{\line(1,0){\nx}}
}

\def\puti#1#2#3{
\nxo=#1
\nyo=#2
\nl=4
\multiply \nxo by \nl
\multiply \nyo by \nl
\divide \nl by 2
\advance \nxo by - \nl
\advance \nxo by - 1
\advance \nyo by - \nl
\advance \nyo by - 1
\put(\nxo,\nyo){$#3$}
}

\def\putl#1#2#3{
\nxo=#1
\nyo=#2
\nl=4
\multiply \nxo by \nl
\multiply \nyo by \nl
\divide \nl by 2
\advance \nxo by - \nl
\advance \nyo by - \nl
\put(\nxo,\nyo){$#3$}
}
\def\puto#1#2{
\nxo=#1
\nyo=#2
\nl=4
\thicklines
\multiply \nxo by \nl
\multiply \nyo by \nl
\divide \nl by 2
\advance \nxo by - \nl
\advance \nyo by - \nl
\put(\nxo,\nyo){\circle{1.5}}
}
\def\case#1#2#3#4{
\nxo=#1
\nyo=#2
\nx=#3
\ny=#4
\nl=4
\multiply \nxo by \nl
\multiply \nyo by \nl
\advance \nxo by - \nl
\advance \nyo by - \nl
\multiply \ny by  \nl
\linethickness{0.3mm}
\multiply \nx by \nl
\multiput(\nxo,\nyo)(\nx,0){2}%
{\line(0,1){\ny}}
\nx=#3
\ny=#4
\multiply \nx by  \nl
\multiply \ny by  \nl
\multiput(\nxo,\nyo)(0,\ny){2}%
{\line(1,0){\nx}}
}
\def\putc#1#2{
\nxo=#1
\nyo=#2
\multiply \nxo by 4
\multiply \nyo by 4
\divide \nl by 2
\advance \nxo by - 2
\advance \nyo by -2
\thicklines
\put(\nxo,\nyo){\circle{1.5}}
\put(\nxo,\nyo){\circle{0.8}}
}
\def\putcc#1#2{
\nxo=#1
\nyo=#2
\nl=4
\multiply \nxo by \nl
\multiply \nyo by \nl
\divide \nl by 2
\advance \nxo by - \nl
\advance \nyo by - \nl
\thicklines
\put(\nxo,\nyo){\circle{2.0}}
\put(\nxo,\nyo){\circle{1.5}}
\put(\nxo,\nyo){\circle{0.8}}
}
\def\const#1#2{
\nxo=#1
\nyo=#2
\multiply \nxo by 4
\multiply \nyo by 4
\advance \nxo by - 2
\advance \nyo by - 2
\put(\nxo,\nyo){\circle*{2.7}}
}

\def\constx#1#2{
\setlength{\unitlength}{0.3mm}
\nxo=#1
\nyo=#2
\linethickness{0.8mm}
\multiply \nxo by 20
\multiply \nyo by 20
\advance \nxo by - 19%
\advance \nyo by - 18%
\put(\nxo,\nyo){\line(1,0){18}}%
\advance \nyo by 16
\put(\nxo,\nyo){\line(1,0){18}}
\advance \nyo by -17
\advance \nxo by 1
\put(\nxo,\nyo){\line(0,1){18}}
\advance \nxo by 16
\put(\nxo,\nyo){\line(0,1){18}}
\setlength{\unitlength}{1.5mm}
}
\long\def\symbolfootnote[#1]#2{\begingroup%
\def\thefootnote{\fnsymbol{footnote}}\footnote[#1]{#2}\endgroup} 

\numberwithin{equation}{section}

\begin{document}
\thispagestyle{empty}
\begin{flushright}
\end{flushright}
\baselineskip=16pt
\vspace{.5in}
{
\begin{center}
{\bf Electric-magnetic Duality of Abelian Gauge Theory on the Four-torus,}

{\bf from the Fivebrane on ${\bf T^2\times T^4}$, via their Partition Functions}
\end{center}}
\vskip 1.1cm
\begin{center}
\centerline{ Louise Dolan\symbolfootnote[1] 
{E-mail: ldolan@physics.unc.edu}
and Yang Sun\symbolfootnote[2]{E-mail: sylmf@email.unc.edu}}
\bigskip
\centerline{\em Department of Physics}
\centerline{\em University of North Carolina, Chapel Hill, NC 27599} 
\bigskip
\vskip5pt

\bigskip
\bigskip
\bigskip
\bigskip
\end{center}

\abstract{\noindent 
We compute the partition function of four-dimensional abelian gauge
theory on a general four-torus $T^4$ with flat metric using 
Dirac quantization. In addition to an $SL(4,{\cal Z})$ symmetry,
it possesses $SL(2,{\cal Z})$ symmetry that is electromagnetic S-duality.
We show explicitly how this $SL(2,{\cal Z})$ S-duality of the 
$4d$ abelian gauge theory
has its origin in symmetries of the $6d$ $(2,0)$ tensor theory,
by computing the partition function of a single
fivebrane compactified on $T^2$ times $T^4$, which has $SL(2,{\cal Z})\times 
SL(4,{\cal Z})$ symmetry. If we identify the couplings of the abelian 
gauge theory $\tau = \frac{\theta}{2\pi} + i\frac{4\pi}{e^2} $ with the 
complex modulus of the $T^2$ torus $\tau = \beta^2 + i \frac{R_1}{R_2}$, 
then in the small $T^2$ limit, the partition function of the fivebrane tensor
field can be factorized, and contains the partition function of the $4d$ gauge
theory. In this way the $SL(2,{\cal Z})$ symmetry of the 6d tensor
partition function
is identified with the $S$-duality symmetry of the 4d gauge partition function.
Each partition function is the product of zero mode and oscillator 
contributions, where the $SL(2,{\cal Z})$ acts suitably. 
For the 4d gauge theory, which has a Lagrangian,
this product redistributes when using path integral quantization.
\bigskip}
\thispagestyle{empty}
\setlength{\parindent}{0pt}
\setlength{\parskip}{6pt}

\setstretch{1.05}
\vfill\eject
\vskip50pt

\section{Introduction}
\label{Introduction}
Four-dimensional $N=4$ Yang-Mills theory is conjectured 
to possess $S$-duality,
which implies 
the theory with gauge coupling $g$, gauge group $G$, and theta parameter
$\theta$ is equivalent 
to one with $\tau\equiv {\theta\over 2\pi} + {4\pi i\over g^2}$ transformed
by modular transformations $SL(2,{\cal Z})$, 
and the group to $G^{\vee}$ \cite{MO}-\cite{WO}, with the 
weight lattice of $G^{\vee}$ dual to that of $G$ . 
The conjecture has been tested by the Vafa-Witten partition 
function on various four-manifolds \cite{VafaWitten}. 
More recently, a computation of 
the $N=4$ Yang-Mills partition function on the four-sphere using the 
localization method for quantization, 
enables checking S-duality directly \cite{Pestun}.

This duality is believed to have its origin in a certain
superconformal field theory 
in six dimensions, the M5 brane $(2,0)$ theory.
When the $6d$, $N=(2,0)$ theory is compactified on $T^2$, 
one obtains the $4d$, $N=4$ Yang-Mills theory, 
and the $SL(2, {\cal Z})$ group of the torus should imply the 
S-duality of the four-dimensional gauge theory 
\cite{Witten2}-\cite{Verlinde}. 

In this paper, we compare 
the partition function of the $6d$ chiral tensor boson of one fivebrane 
compactified on $T^2 \times T^4$, with that of
$U(1)$ gauge theory with a $\theta$ parameter, compactified on $T^4$.
We use these to 
show explicitly how the $6d$ theory is the origin of S-duality in 
the gauge theory.
Since the $6d$ chiral boson has a self-dual three-form field strength
and thus lacks a Lagrangian \cite{Witten3}, we will use the 
Hamiltonian formulation to
compute the partition functions for both theories. 

As motivated by \cite{GSW}, 
the four-dimensional $U(1)$ gauge partition function on $T^4$ is
\begin{align} Z^{4d, Maxwell}&\equiv
tr e^{-2\pi H^{4d} + i 2\pi \gamma^\alpha P^{4d}_\alpha} =
Z^{4d}_{\rm zero\;modes}\;\cdot Z^{4d}_{\rm osc},\label{4dpf}\end{align}
where the Hamiltonian and momentum are
\begin{align}
{H}^{4d} = \int_0^{2\pi} d^3\theta\Big(
\frac{e^2}{4} \frac{R^2_6}{\sqrt{g}} 
g_{\alpha\beta} \Pi^{\alpha} \Pi^{\beta}  &+ \frac{e^2}{32\pi^2} 
\sqrt{g} \left[\frac{\theta^2}{4\pi^2} + \frac{16\pi^2}{e^4}\right] 
g^{\alpha\beta} g^{\gamma\delta} F_{\alpha\gamma} F_{\beta\delta}
+ \frac{\theta e^2}{16\pi^2} \frac{R^2_6}{\sqrt{g}} 
g_{\alpha\beta}
\epsilon^{\alpha\gamma\delta}F_{\gamma\delta}\Pi^{\beta}\Big),\cr
P^{4d}_{\alpha} 
&= \int_0^{2\pi} d^3 \theta \; \Pi^\beta
F_{\alpha\beta}\label{wholepf}\end{align}
in terms of the gauge field strength tensor $F_{ij} 
(\theta^3, \theta^4, \theta^5,\theta^6)$, the conjugate momentum $\Pi^\alpha$,
and the constant parameters
$g_{\alpha\beta}, R_6$ and $\gamma^{\alpha}$ in the metric $G_{ij}$ of $T^4$. 
They will be derived from the abelian gauge theory Lagrangian, given here
for Euclidean signature
\begin{align}
I = \frac{1}{8\pi} \int_{T^4} d\theta_3 d\theta_4 d\theta_5 d\theta_6 
\,\big(\frac{4\pi}{e^2} \sqrt{g} F^{ij} F_{ij} - \frac{i\theta}{4\pi} 
\epsilon^{ijkl} F_{ij} F_{kl} \big),
\label{actiontheta2}\end{align}
with $\epsilon^{3456}=1$, $\epsilon_{ijkl} = g \epsilon^{ijkl}$, 
and $g = {\rm det} (G_{ij})$. 

In contrast, the partition function of the 
abelian chiral two-form on $T^2\times T^4$ is \cite{DN}
\begin{align}
Z^{6d, chiral} &=tr \,e^{-2\pi R_6{\cal H} + i2\pi\gamma^\alpha{\cal P}_\alpha}
= Z^{6d}_{\rm zero \; modes}  \cdot Z^{6d}_{\rm osc},\cr
{\cal H} &= {1\over 12}\int_0^{2\pi} d\theta^1 \ldots d\theta^5 
{\sqrt G_5}{G_5}^{mm'}
{G_5}^{nn'}{G_5}^{pp'} H_{mnp}(\vec\theta,\theta^6)\,
H_{m'n'p'}(\vec\theta,\theta^6),\cr
{\cal P}_\alpha&= -{1\over 24}{\int_0}^{2\pi}
d\theta^1...d\theta^5 \epsilon^{mnprs} H_{mnp}(\vec\theta,\theta^6)\,
H_{\alpha rs}(\vec\theta,\theta^6)
\label{6dpf}\end{align}
where $\theta^1$ and $\theta^2$ are the coordinates of the two one-cycles of 
$T^2$. The time direction $\theta^6$ is common to both theories, the 
angle between $\theta^1$ and $\theta^2$ is $\beta^2$, and ${G_5}^{mn}$ is the 
inverse metric of ${G_5}_{mn}$, where $1\leq m,n \leq 5$. 
The eight angles between the two-torus and the four-torus are set to 
zero.\footnote{A different consideration of the fivebrane on $T^2\times T^4$ in 
\cite{BG} includes the time direction in $T^2$.}

Section 2 is a list of our results; their derivations are presented 
in the succeeding sections.
In section 3, the contribution of the zero modes to the partition function 
for the chiral theory on the manifold 
$M = T^2 \times T^4$ is computed as a sum over 
ten integer eigenvalues using the Hamiltonian formulation. 
The zero mode sum for the gauge theory on the same $T^4 \subset M$ 
is calculated with six integer eigenvalues. 
We find that once we identify the modulus of the $T^2$
contained in $M$, 
$\tau = \beta^2 + i \frac{R_1}{R_2}$, with the gauge couplings 
$\tau = \frac{\theta}{2\pi} + i \frac{4\pi}{e^2}$, then the two theories are 
related by $Z^{6d}_{zero\,modes} = \epsilon Z^{4d}_{zero\,modes }$, 
where $\epsilon$ is due to the zero modes of the scalar field that 
arises in addition to $F_{ij}$
from the compactification of the $6d$ self-dual three-form.  
In section 4, the abelian gauge theory is quantized on a four-torus using 
Dirac constraints, and the Hamiltonian and momentum are computed in terms 
of oscillator modes. 
For small $T^2$, the Kaluza-Klein modes are removed from 
the partition function of the chiral two-form, and in this limit 
it agrees with the gauge theory result, up to the scalar field 
contribution. 
In Appendix A, we show 
the path integral quantization gives the same result for the 4d gauge theory
partition function as canonical quantization. However, 
the zero and oscillator mode contributions differ in the two quantizations.
In Appendix B, we show how the zero and oscillator mode contributions 
transform under $SL(2, {\cal Z})$ for the 6d theory, 
as well as for both quantizations
of the 4d theory. We prove the partition functions in 4d and 6d
are both $SL(2, {\cal Z})$ invariant.
In Appendix C, the vacuum energy is regularized. 
In Appendix D, we introduce a complete set of 
$SL(4, {\cal Z})$ generators, and then prove the 4d and 6d partition functions 
are invariant under $SL(4, {\cal Z})$ transformations. 
\vfill\eject

\section{\bf Statement of the main result}

We compute partition functions for a chiral boson on $T^2 \times T^4 $ 
and for a $U(1)$ gauge boson on the same $T^4$. 
The geometry of the manifold $T^2 \times T^4$ will be described by the 
line element,
\begin{align}
ds^2 &= R_2^2 (d\theta^2 - \beta^2 d\theta^1)^2 + R_1^2 (d\theta^1)^2 \cr 
&+ \sum_{\alpha,\beta} g_{\alpha\beta}(d\theta^\alpha - \gamma^{\alpha} 
d\theta^6) (d\theta^\beta - \gamma^{\beta} d\theta^6) + R_6^2 (d\theta^6)^2,
\label{lineelem}\end{align}
with $0 \leq \theta^I \leq 2\pi$, $1 \leq I \leq 6$, and 
$3 \leq \alpha \leq 5$. $R_1$, $R_2$ are the radii for 
directions $I=1, 2$ on $T^2$, and $\beta^2$ is the angle between them.  
$g_{\alpha\beta}$ fixes the metric for a $T^3$ submanifold of $T^4$, 
$R_6$ is the remaining radius, and $\gamma^{\alpha}$ is the angle between 
those. So, from (\ref{lineelem}) the metric is
\begin{align}
T^2: \qquad &G_{11} = {R_1}^2 + R_2^2 \beta^2 \beta^2, \qquad G_{12} 
= - R_2^2 \beta^2, \qquad G_{22} = R_2^2;\cr
T^4: \qquad &G_{\alpha \beta} = g_{\alpha\beta}, 
\qquad G_{\alpha 6} = -g_{\alpha\beta}\gamma^{\beta}, \qquad G_{66} ={R_6}^2 + g_{\alpha\beta} \gamma^{\alpha}\gamma^{\beta};\cr
&G_{\alpha 1} = G_{\alpha 2} = 0, \hskip30pt G_{16} = G_{26} =0;
\label{metric1}\end{align} and the inverse metric is
\begin{align}
T^2: \qquad &G^{11} = \frac{1}{{R_1}^2},\qquad G^{12} = \frac{\beta^2}{{R_1}^2}, \qquad G^{22} = \frac{1}{R_2^2} + \frac{\beta^2\beta^2}{{R_1}^2} \equiv g^{22} +\frac{\beta^2\beta^2}{{R_1}^2}  ;\cr
T^4: \qquad &G^{\alpha\beta} = g^{\alpha\beta} 
+ \frac{\gamma^{\alpha}\gamma^{\beta}}{{R_6}^2},\qquad G^{\alpha 6} 
= \frac{\gamma^{\alpha}}{{R_6}^2},\qquad G^{66} = \frac{1}{{R_6}^2}\cr 
&G^{1\alpha} = G^{2\alpha} = 0, \hskip30pt G^{16} = G^{26} = 0.
\label{6dinverse}\end{align} 
$\theta^6$ is chosen to be the time direction for both theories. In the $4d$ 
expression (\ref{actiontheta2}) the indices of the field strength tensor 
have $3 \leq i, j, k, l \leq 6$, whereas in (\ref{6dpf}), 
the Hamiltonian and momentum are written in terms of fields with indices 
$1 \leq m, n, p, r, s \leq 5$. 
The 5-dimensional inverse in directions 1, 2, 3, 4, 5  is ${G_5}^{mn}$, 
\begin{align}
&G_5^{11} = \frac{1}{{R_1}^2},\qquad G_5^{12} = \frac{\beta^2}{{R_1}^2},
\qquad G_5^{22} = g^{22} + \frac{\beta^2\beta^2}{{R_1}^2}\cr
&G_5^{\alpha\beta} = g^{\alpha\beta},\qquad G_5^{1\alpha} = 0, 
\qquad G_5^{2\alpha} = 0.
\label{metric2}\end{align} 
$g^{\alpha\beta}$ is the $3d$ inverse of $g_{\alpha\beta}.$
The determinants are related by 
\begin{align}\sqrt{G} =  \sqrt{{\rm det} G_{IJ} }  = R_1 R_2 \sqrt{g} 
= R_1 R_2 R_6 \sqrt{\tilde g} = R_6 \sqrt{G_5},\label{determinantmetric}
\end{align} where $G$ is the determinant for $6d$ metric $G_{IJ}$.$\;$
$G_5$, $g$ and ${\tilde g}$ are the determinants for the 
$5d$ metric ${G}_{mn}$, 
$4d$ metric $G_{ij}$, and $3d$ metric $g_{\alpha\beta}$ respectively. 

The zero mode partition function of the $6d$ 
chiral boson on $T^2 \times T^4$ with the metric (\ref{6dinverse}) is

\begin{align}
Z^{6d}_{\rm zero\, modes} =& \sum_{n_8,n_9,n_{10}} {\rm exp} 
\{-
\frac{\pi R_6}{R_1 R_2}\sqrt{\tilde g} 
g^{\alpha\alpha'}H_{12\alpha}H_{12\alpha'} \}\cr 
&\hskip10pt
\cdot \sum_{n_7}{\rm exp} \{-\frac{\pi}{6}R_6R_1 R_2 \sqrt{\tilde g}
g^{\alpha\alpha'}g^{\beta\beta'}g^{\delta\delta'}H_{\alpha\beta\delta}H_{\alpha'\beta'\delta'}  -i\pi\gamma^{\alpha}\epsilon^{\gamma\beta\delta}H_{12\gamma}H_{\alpha\beta\delta}  \}\cr
&\cdot \hskip-6pt
\sum_{n_4,n_5,n_6} {\rm exp} \{-\frac{\pi}{2}R_6R_1 R_2\sqrt{\tilde g}(\frac{1}{R_2^2}+\frac{{\beta^2}^2}{{R_1}^2})g^{\alpha\alpha'}g^{\beta\beta'}H_{2\alpha\beta}H_{2\alpha'\beta'} \}\cr
&\hskip10pt\cdot\hskip-6pt
\sum_{n_1,n_2,n_3} {\rm exp} \{-\pi\frac{R_6 R_2}{R_1} \sqrt{\tilde g} 
\beta^2 g^{\alpha\alpha'}g^{\beta\beta'}H_{1\alpha\beta}H_{2\alpha'\beta'} 
+ i\pi\gamma^{\alpha}\epsilon^{\gamma\beta\delta} H_{1\gamma\beta} 
H_{2\alpha\delta}\cr 
&\hskip70pt
-\frac{\pi}{4}\frac{R_6 R_2}{R_1} \sqrt{\tilde g}(g^{\alpha\alpha'}g^{\beta\beta'}-g^{\alpha\beta'}g^{\beta\alpha'})H_{1\alpha\beta}H_{1\alpha'\beta'} \}
\label{zmpf}\end{align}
where the zero mode eigenvalues of the field strength tensor are integers,
and (\ref{zmpf}) factors into a sum on
$H_{\alpha\beta\gamma}$ as $H_{345} = n_7$,
$H_{12\alpha}$ as $H_{123} = n_8$,
$H_{124} = n_9$, $H_{125} = n_{10}$; and a sum over
$H_{1\alpha\beta}$ defined as $H_{134} = n_1$, 
$H_{145} = n_2$, $H_{135} = n_3$ and $H_{2\alpha\beta}$ as $H_{234} = n_4$, 
$H_{245} = n_5$, $H_{235} = n_6$, as we will show in section 3.

The zero mode partition function of the $4d$ gauge boson on $T^4$ 
with the metric (\ref{metric1}) is 
\begin{align}
&Z^{4d}_{\rm{zero\, modes}}= \sum_{n_4, n_5, n_6}  {\rm exp}
\{ -{e^2\over 4} {R^2_6\over\sqrt{g}}
g_{\alpha\beta} {\widetilde \Pi}^{\alpha}
{\widetilde \Pi}^{\beta}\}
\cdot \sum_{n_1, n_2, n_3} {\rm exp}
\{-{\theta e^2\over 8\pi}
\frac{R^2_6}{\sqrt{g}} g_{\alpha\beta}
\epsilon^{\alpha\gamma\delta} {\widetilde F}_{\gamma\delta}
{\widetilde \Pi}^{\beta} \}\cr
&\hskip110pt\cdot {\rm exp} \{- \frac{e^2\sqrt{g}}{8}
({\theta^2\over 4\pi^2} + {16\pi^2\over e^4}) g^{\alpha\beta}
g^{\gamma\delta} {\widetilde F}_{\alpha\gamma} {\widetilde F}_{\beta\delta}
+ 2\pi i \gamma^\alpha
{\widetilde \Pi}^{\beta} {\widetilde F}_{\alpha\beta} \},
\label{zmpf4d}\end{align}
where $\widetilde \Pi^\alpha$ take integer values 
$\widetilde \Pi^3=n_4,\, 
\widetilde \Pi^4=n_5, \,\widetilde \Pi^5=n_6,$ and $\widetilde F_{34}=n_1,$ 
$\widetilde F_{35}=n_2, \,\widetilde F_{45}=n_3$, from section 3.  
We identify the integers
\begin{align}
H_{2\alpha\beta} = \widetilde F_{\alpha\beta}  \hskip50pt  \hbox{and} 
\hskip50pt H_{1\alpha\beta} = \frac{1}{\tilde g} \epsilon_{\alpha\beta\gamma}
{\widetilde \Pi}^{\gamma},
\end{align}
where $\tilde g=g\,R^{-2}_6$ from (\ref{determinantmetric}),
and the modulus $$\tau =  \beta^2 + i\frac{R_1}{R_2} = 
\frac{\theta}{2\pi} + i\frac{4\pi}{e^2},$$
so that as shown in section 3,
we have the factorization
\begin{align}
Z^{6d}_{\rm zero\, modes} &= \epsilon \, Z^{4d}_{\rm zero\, modes},
\label{zmpfsep}\end{align} where $\epsilon$ comes from
the remaining four zero modes 
$H_{\alpha\beta\gamma}$ and $H_{12\alpha}$ due to the additional scalar 
that occurs in the compactification of the $6d$ self-dual three-form
field strength,

\begin{align}
\epsilon =&\sum_{n_8,n_9,n_{10}} {\rm exp}\{ -
\frac{\pi R_6}{R_1 R_2}\sqrt{\tilde g} g^{\alpha\alpha'}
H_{12\alpha}H_{12\alpha'} \}\cr 
&\cdot \sum_{n_7}{\rm exp} \{-\frac{\pi}{6}R_6R_1 R_2 
\sqrt{\tilde g}g^{\alpha\alpha'}g^{\beta\beta'}g^{\delta\delta'}
H_{\alpha\beta\delta}H_{\alpha'\beta'\delta'}  
-i\pi\gamma^{\alpha}\epsilon^{\gamma\beta\delta}
H_{12\gamma}H_{\alpha\beta\delta}\}.
\label{epsil}\end{align}
From section 4, 
there is a similar relation between the oscillator partition functions
\begin{align}
{\rm lim}_{R_1, R_2 \rightarrow 0} Z^{6d}_{\rm osc } &= 
\epsilon' Z^{4d}_{\rm osc},
\label{osclim}\end{align}
where
\begin{align}
Z^{6d}_{\rm osc} &= 
\Bigl ( e^{ R_6 \pi^{-3} \sum_{\vec n\ne \vec 0} {\sqrt{G_5}\over
(G_{mp} n^mn^p)^{\,3}}} \prod_{\vec p\in {\cal Z}^5\ne \vec 0}
{\textstyle 1\over{1- e^{-2\pi R_6 \sqrt{g^{\alpha\beta} p_\alpha p_\beta
+\tilde p^2} + 2\pi i\gamma^\alpha p_\alpha}}}\Bigr )^3,
\label{f6f}\end{align}
\begin{align}
Z^{4d}_{\rm osc}&= \Big( e^{{1 \over 2}R_6\pi^{-2}
\sum_{\vec n\ne 0} {\sqrt{\tilde g}\over 
({g_{\alpha\beta}n^\alpha n^\beta})^{2}}}
\; \cdot \prod_{\vec n\in {\cal Z}^3\ne \vec 0} {1\over 1 - e^{-2\pi
R_6\sqrt{g^{\alpha\beta} n_\alpha n_\beta} 
- 2\pi i \gamma^\alpha n_\alpha}} \Big)^2,
\label{f4f}\end{align}
where
${\tilde p}^2\equiv {p_1^2\over R_1^2} + ({1\over R_1^2} + {\beta^2\beta^2\over 
R_2^2}) p_2^2 + {2\beta^2\over R_1^2} p_1p_2,$
and $\epsilon'$ is the oscillator contribution from the additional scalar,
\begin{align}
\epsilon' =  e^{{1 \over 2}R_6\pi^{-2}
\sum_{\vec n\ne 0} {\sqrt{\tilde g}\over 
({g_{\alpha\beta}n^\alpha n^\beta})^{2}}}
\; \cdot \prod_{\vec n\in {\cal Z}^3\ne \vec 0} {1\over 1 - e^{-2\pi
R_6\sqrt{g^{\alpha\beta} n_\alpha n_\beta} - 2\pi i \gamma^\alpha n_\alpha}}.
\end{align}
Therefore, in the limit of small $T^2$, we have
\begin{align}
{\rm lim}_{R_1, R_2 \rightarrow 0} \, Z^{6d,\,chiral} 
= \epsilon \, \epsilon' \, Z^{4d,\,Maxwell}.
\label{llim4d}\end{align}
We use this relation between the $6d$ and $4d$ partition functions
to extract the S-duality of the latter from a geometric symmetry of the
former.
For $\tau = \beta^2 + i \frac{R_1}{R_2} = \frac{\theta}{2\pi} 
+ i \frac{4\pi}{e^2}$, under the $SL(2,{\cal Z})$ transformations
\begin{align}
\tau \rightarrow -\frac{1}{\tau}; \quad {\tau} \rightarrow \tau -1,
\end{align}
$Z^{6d}_{\rm{zero\,modes}}$ and $Z^{6d}_{\rm{osc}}$ are separately invariant, 
as are $Z^{4d}_{\rm{zero \,modes}}$ and $Z^{4d}_{\rm{osc}}$, which  
we will prove in Appendix B. In particular, $Z^{4d}_{\rm{osc}}$
is independent of $e^2$ and $\theta$.
A path integral computation agrees with our $U(1)$ partition function,
as we review in Appendix A \cite{Zucchini}. 
Nevertheless, in the path integral quantization the zero and non-zero mode
contributions are rearranged, and although each is invariant under
$\tau\rightarrow\tau-1$, they transform differently under $\tau
\rightarrow -{1\over \tau}$, with $Z^{PI}_{\rm zero\, modes}
\rightarrow |\tau|^3 Z^{PI}_{\rm zero\, modes}$ and
$Z^{PI}_{\rm non-zero\, modes}
\rightarrow |\tau|^{-3} Z^{PI}_{\rm non-zero\, modes}.$
For a general spin manifold, 
the $U(1)$ partition function transforms
as a modular form under S-duality \cite{Witten0}, but in the case of $T^4$
which we consider in this paper the weight is zero. 

\section{\bf Zero Modes}
\label{Zero Modes}
In this section, we show details for the computation of the 
zero mode partition functions.
The $N = (2,0)$,  $6d$ world volume theory of the fivebrane contains a chiral 
two-form $B_{MN}$, which has a self-dual three-form field strength 
$H_{LMN} = \partial_L B_{MN} + \partial_{M} B_{NL} + \partial_N B_{LM}$ 
with $1 \leq L, M, N \leq 6$,
\begin{align}
H_{LMN} (\vec \theta, \theta^6) = \frac{1} {6\sqrt{-G}} G_{LL'} G_{MM'} G_{NN'} \epsilon^{L'M'N'RST} H_{RST} (\vec \theta, \theta^6).
\label{fieldstrengthtwoform}\end{align}
Since there is no covariant Lagrangian description for the chiral two-form,  
we compute its partition function from (\ref{6dpf}).  
As in \cite{DN},\cite{DS}  
the zero mode partition function of the $6d$ chiral theory is calculated 
in the Hamiltonian formulation similarly to string theory,
\begin{align}
{Z}^{6d}_{\rm zero \; modes} = {\it tr} \big( e^{-t{\cal H} + iy^l {\cal P}_l} 
\big)
\label{partition}\end{align}
where $t=2\pi R_6$ and $y^l = 2\pi \frac{G^{l6}}{G^{66}}$, with $l=1,..5$. 
However, $y^1$ and $y^2$ are zero due to the metric (\ref{6dinverse}). 
Neglecting the integrations and using the metric (\ref{metric2}) in
(\ref{6dpf}), we find 
\begin{align}
-t{\cal H} &= 
-\frac{\pi}{6} R_6 R_1 R_2 \sqrt{\tilde g} g^{\alpha\alpha'} g^{\beta\beta'} g^{\lambda\lambda'} H_{\alpha\beta\lambda} H_{\alpha'\beta'\lambda'} - 
\frac{\pi}{2} R_6  \frac{R_1}{R_2} \sqrt{\tilde g} g^{\alpha\alpha'}g^{\beta\beta'} H_{2\alpha\beta}H_{2\alpha'\beta'} \cr &-\frac{\pi}{2} \frac{R_6}{R_1} R_2 {\beta^2}^2 g^{\alpha\alpha'} g^{\beta\beta'} \sqrt{\tilde g} H_{2\alpha\beta} H_{2\alpha'\beta'}-\pi\frac{R_6}{R_1} R_2\beta^2 \sqrt{\tilde g} g^{\alpha\alpha'}g^{\beta\beta'} H_{1\alpha\beta} H_{2\alpha'\beta'}\cr
&-\frac{\pi R_6}{R_1 R_2} \sqrt{\tilde g} g^{\alpha\alpha'}  
H_{12\alpha} H_{12\alpha'} - \frac{\pi}{4} R_2 \frac{R_6}{R_1} \sqrt{\tilde g}
(g^{\alpha\alpha'}g^{\beta\beta'} - g^{\alpha\beta'}g^{\alpha'\beta})
H_{1\alpha\beta}H_{1\alpha'\beta'},
\label{6dHam}\end{align}
and the momentum components $3 \leq \alpha \leq 5$ are 
\begin{align}
{\cal P}_{\alpha} &
= -\frac{1}{2} \epsilon^{\gamma\beta\delta}H_{12\gamma}H_{\alpha\beta\delta} 
+\frac{1}{2} \epsilon^{\gamma\beta\delta}H_{1\gamma\beta}H_{2\alpha\delta},
\end{align}
where the zero modes of the ten fields $H_{lmp}$ are labeled by integers
$n_1,\ldots n_{10}$ \cite{DN}.
Then (\ref{partition}) is given by (\ref{zmpf}).
\vskip10pt

Similarly, we compute the zero mode partition function for the 4d $U(1)$ 
theory from (\ref{4dpf}). 
We consider the charge quantization condition
\begin{align}
&n_I = {1\over 2\pi}\int_{\Sigma_2^I} F\equiv  {1\over 2\pi}\int_{\Sigma_2^I}
{1\over 2} F_{\alpha\beta} \, d\theta^\alpha\wedge d\theta^\beta,\qquad n_I\in {\cal Z},
\; \hbox{for each} \; 1\le I\le 3.\label{coho}
\end{align}
as well as the commutation relation obtained from (\ref{thcr})
\begin{align}
\left[\int_{\Sigma_1^\gamma} A_\alpha(\vec\theta,\theta^6) d\theta^\alpha,
\int {d^3\theta'\over 2\pi}  \Pi^\beta(\vec\theta',\theta^6)\right]
= {i\over 2\pi} \int_{\Sigma_1^\gamma} d\theta^\beta = i \,\delta_\gamma^\beta,
\end{align}
and use the standard argument \cite{DS},\cite{Henningsonone} to show that
the field strength $F_{\alpha\beta}$ and momentum $\Pi^\alpha$ zero modes
have eigenvalues
\begin{align}
F_{\alpha\beta} = {n_{\alpha,\beta}\over 2\pi}, \quad n_{\alpha, \beta}\in {\cal Z} \; \hbox{for $\alpha<\beta$}, \; \qquad \;\hbox{and} \; \qquad \Pi^\alpha(\vec\theta',\theta^6)
= {n^{(\alpha)}\over (2\pi)^2},\quad n^{(\alpha)}\in {\cal Z}^3.
\label{Fint}\end{align} 
Thus we define integer valued modes $\widetilde F_{\alpha\beta}
\equiv 2\pi F_{\alpha\beta}$ and
${\widetilde \Pi}^{\alpha} \equiv (2\pi)^2 \Pi^{\alpha}$.
Taking into account the spatial integrations $d\theta^\alpha$, 
(\ref{wholepf}) gives
\begin{align}
&- 2\pi H^{4d} + i 2\pi \gamma^\alpha P_\alpha^{4d}\cr
&= -\frac{e^2}{4} \frac{R^2_6}{\sqrt{g}} g_{\alpha\beta} 
{\widetilde \Pi}^{\alpha} {\widetilde \Pi}^{\beta}  
- \frac{e^2\sqrt{g}}{8} \left[\frac{\theta^2}{4\pi^2} 
+ \frac{16\pi^2}{e^4}\right] g^{\alpha\beta} g^{\gamma\delta} 
{\widetilde F}_{\alpha\gamma} {\widetilde F}_{\beta\delta} 
- \frac{\theta e^2}{8\pi} \frac{R^2_6}{\sqrt{g}} g_{\alpha\beta} 
\epsilon^{\alpha\gamma\delta} {\widetilde F}_{\gamma\delta} 
{\widetilde \Pi}^{\beta}\cr
&\hskip10pt
+2\pi i \gamma^\alpha
{\widetilde \Pi}^{\beta} {\widetilde F}_{\alpha\beta},
\label{maxwell4d}\end{align}
where (\ref{wholepf}) itself is derived in section 4.
So from (\ref{maxwell4d}) and (\ref{4dpf}), 
\begin{align}
&Z^{4d}_{\rm{zero\, modes}}= \sum_{n_4, n_5, n_6}  {\rm exp}
\{ -{e^2\over 4} {R^2_6\over\sqrt{g}} 
g_{\alpha\beta} {\widetilde \Pi}^{\alpha}
{\widetilde \Pi}^{\beta}\}  
\cdot \sum_{n_1, n_2, n_3} {\rm exp}
\{-{\theta e^2\over 8\pi} 
\frac{R^2_6}{\sqrt{g}} g_{\alpha\beta}
\epsilon^{\alpha\gamma\delta} {\widetilde F}_{\gamma\delta}
{\widetilde \Pi}^{\beta} \}\cr
&\hskip110pt\cdot {\rm exp} \{- \frac{e^2\sqrt{g}}{8} 
({\theta^2\over 4\pi^2} + {16\pi^2\over e^4}) g^{\alpha\beta}
g^{\gamma\delta} {\widetilde F}_{\alpha\gamma} {\widetilde F}_{\beta\delta}
+ 2\pi i \gamma^\alpha
{\widetilde \Pi}^{\beta} {\widetilde F}_{\alpha\beta} \},
\label{zmpfagain}\end{align}
where $n_I$ are integers, with
 $\widetilde F_{34}=n_1,$
$\widetilde F_{35}=n_2, \,\widetilde F_{45}=n_3$, and
$\widetilde \Pi^3=n_4,\,
\widetilde \Pi^4=n_5, \,\widetilde \Pi^6=n_6.$
(\ref{zmpfagain}) is 
the zero mode contribution to the $4d$ $U(1)$
partition function (\ref{4dpf}), and is (\ref{zmpf4d}).

If we identify the gauge couplings
$\tau = \frac{\theta}{2\pi} + i\frac{4\pi}{e^2}$ with 
the modulus of $T^2$, $\tau = \beta^2 + i \frac{R_1}{R_2}$, then
\begin{align}
\frac{e^2}{4\pi} = \frac{R_2}{R_1}, \qquad  \frac{\theta}{2\pi} = \beta^2,
\label{taupara}\end{align}
and (\ref{zmpfagain}) becomes
\begin{align}
&Z^{4d}_{\rm{zero\, modes}}= \sum_{n_4, n_5, n_6}  {\rm exp}
\{ -\pi \frac{R_2R^2_6}
{R_1\sqrt{g}} g_{\alpha\beta} {\widetilde \Pi}^{\alpha}
{\widetilde \Pi}^{\beta}\}  \cdot \sum_{n_1, n_2, n_3} {\rm exp}
\{- \pi \beta^2
\frac{R_2 R^2_6}{ R_1 \sqrt{g}} g_{\alpha\beta}
\epsilon^{\alpha\gamma\delta} {\widetilde F}_{\gamma\delta}
{\widetilde \Pi}^{\beta} \}\cr
&\hskip110pt\cdot {\rm exp} \{- \frac{\pi}{2} \frac{R_2}{R_1}
\sqrt{g} ({\beta^2}^2 + \frac{R_1^2}{R_2^2}) g^{\alpha\beta}
g^{\gamma\delta} {\widetilde F}_{\alpha\gamma} {\widetilde F}_{\beta\delta}
+ 2\pi i \gamma^\alpha
{\widetilde \Pi}^{\beta} {\widetilde F}_{\alpha\beta} \}.
\label{zmpfcb}\end{align}
Then the last four terms in the chiral boson
zero mode sum (\ref{zmpf}) are equal to 
(\ref{zmpf4d}) since 
\begin{align}
-\frac{\pi}{2} \frac{R_2 R_6 \sqrt{\tilde g}}{R_1} (\frac{R_1^2}{R_2^2}
+ {\beta^2}^2 ) g^{\alpha\alpha'} g^{\beta\beta'} H_{2\alpha\beta}
H_{2\alpha'\beta'} &=
-\frac{\pi}{2} \frac{R_2 R_6}{R_1} \sqrt{\tilde g} \,
(\frac{R_1^2}{R_2^2}
+ {\beta^2}^2 ) g^{\alpha\beta} g^{\gamma\delta} \widetilde F_{\alpha\gamma}
\widetilde F_{\beta\delta},\cr
-\pi \frac{R_6 R_2}{R_1} \sqrt{\tilde g} \,\beta^2  g^{\alpha\alpha'}
g^{\beta\beta'} H_{1\alpha\beta} H_{2\alpha'\beta'} &=
-\pi \beta^2\frac{R_6 R_2}{R_1 \sqrt{\tilde g}} g_{\alpha\beta}
\epsilon^{\alpha\gamma\delta} \tilde F_{\gamma\delta} \tilde \Pi^\beta,\cr
i\pi \gamma^{\alpha} \epsilon^{\gamma\beta\delta} 
H_{1\gamma\beta} H_{2\alpha\delta}
&= 2\pi i \gamma^\alpha\,{\tilde \Pi}^\beta {\tilde F}_{\alpha\beta},\cr
-\frac{\pi}{2} \frac{R_6 R_2}{R_1} \sqrt{\tilde g} g^{\alpha\alpha'}
g^{\beta\beta'} H_{1\alpha\beta} H_{1\alpha'\beta'} &= -\pi \frac{R_6 R_2}
{R_1 \sqrt{\tilde g}} g_{\alpha\beta} {\tilde \Pi}^{\alpha}
{\tilde \Pi}^{\beta}, 
\end{align}
when we identify the integers
\begin{align}
H_{2\alpha\beta} = \widetilde F_{\alpha\beta}  \hskip50pt  \hbox{and}
\hskip50pt H_{1\alpha\beta} = \frac{1}{\tilde g} \epsilon_{\alpha\beta\gamma}
{\widetilde \Pi}^{\gamma},
\end{align}
with $\tilde g=g\,R^{-2}_6$ from (\ref{determinantmetric}).
Thus the $6d$ and $4d$ zero mode sums from (\ref{zmpf})  
and (\ref{zmpf4d}) are related by
\begin{align}
Z^{6d}_{\rm zero\, modes} &= \epsilon \,Z^{4d}_{\rm zero\, modes},
\label{zmpfsep3}\end{align} where 
\begin{align}
\epsilon =&\sum_{n_8,n_9,n_{10}} {\rm exp} 
\{-\frac{\pi R_6}{R_1 R_2}\sqrt{\tilde g} g^{\alpha\alpha'}
H_{12\alpha}H_{12\alpha'} \}\cr
&\cdot \sum_{n_7}{\rm exp} \{-\frac{\pi}{6}R_6R_1 R_2
\sqrt{\tilde g}g^{\alpha\alpha'}g^{\beta\beta'}g^{\delta\delta'}
H_{\alpha\beta\delta}H_{\alpha'\beta'\delta'}
-i\pi\gamma^{\alpha}\epsilon^{\gamma\beta\delta}
H_{12\gamma}H_{\alpha\beta\delta}\}.
\end{align}

\section{Oscillator modes}
\label{Oscillator Modes}
To compute the oscillator contribution to the partition function
(\ref{4dpf}), 
we quantize the $U(1)$ gauge theory with a theta term on the $T^4$ 
manifold using Dirac brackets.
From (\ref{actiontheta2}), the equations of motion are
$\partial^i F_{ij} =0$, since the
theta term is a total divergence and does not contribute to them.
So in Lorenz gauge,
the gauge potential $A_i$ with field strength tensor 
$F_{ij} = \partial_i A_j -\partial_j A_i$ 
is obtained by solving the equation
\begin{align}
\partial^{i}\partial_{i} A_{j} = 0,\qquad \hbox{with} \hskip20pt
\partial^{i} A_{i} = 0.
\label{eomagt}\end{align}
The potential has a plane wave solution
\begin{align}
A_{i}(\vec\theta,\theta^6) = \hbox{zero modes}\, +
\sum_{\vec k\ne 0} (f_{i}(k) e^{ik\cdot\theta}
+ (f_{i}(k) e^{ik\cdot\theta})^\ast)\label{biga}\end{align}
with momenta satisfying the on shell condition and gauge condition
\begin{align}
{\widetilde G}_L^{i j} k_{i} k_{j} = 0,
\qquad k^{i} f_{i} =0.
\label{4dg}\end{align}
As in \cite{GSW},\cite{DS}  
the Hamiltonian $H^{4d}$ and momentum $P_{\alpha}^{4d}$ are quantized 
with a Lorentzian signature metric that has zero angles with the time 
direction, $\gamma^\alpha =0$. 
So we modify the metric on the four-torus (\ref{metric1}), (\ref{6dinverse}) 
to be
\begin{align}&{\widetilde G}_{L\,\alpha\beta}= g_{\alpha\beta}\,,\quad
\widetilde G_{L\,66}= -{R_6}^2,\quad
\widetilde G_{L\,\alpha6}=0\,\cr
&\widetilde G_L^{\alpha\beta}= g^{\alpha\beta},\quad
\widetilde G_L^{66}=-{1\over R_6^2},\quad
\widetilde G_L^{\alpha 6}=0,\quad \widetilde G_L = \det 
\widetilde G_{L\,ij} = -g.
\label{Lfourmetric}\end{align}
Solving for $k_6$ from
(\ref{4dg}) we find
\begin{align}
k_6= {\sqrt{-\widetilde G_L^{66}}\over \widetilde G_L^{66}}\,
|k|, \label{k6}\end{align}
where $3\le \alpha,\beta\le 5,$ and $|k| \equiv \sqrt{g^{\alpha\beta} 
k_\alpha k_\beta}$. Employ the remaining gauge invariance\hfill\break
$f_{i}\rightarrow f'_{i} = f_{i} + k_{i} \lambda$ to fix $f'_6=0,$
which is the gauge choice $$A_6=0.$$
This reduces the number of components of $A_i$ from 4 to 3.
To satisfy (\ref{4dg}), we can use the $\partial^{i} F_{i 6}
= -\partial_6\partial^\alpha A_\alpha =0$
component of the equation of motion to eliminate $f_5$ in terms of 
$f_3,f_4$,
\begin{align}
f_5= -{1\over p^5}(p^3f_3+p^4f_4),
\nonumber\end{align}
leaving just two independent polarization vectors
corresponding to the physical degrees of freedom of a four-dimensional 
gauge theory. 

From the Lorentzian Lagrangian and energy-momentum tensor given by
\begin{align}
&{\cal  L} = -\frac{1}{2 e^2} \sqrt{-\widetilde G_L} \widetilde G_L^{ik} 
\widetilde G_L^{jl} 
F_{ij} F_{kl} 
+ \frac{\theta}{32\pi^2} \epsilon^{ijkl} F_{ij} F_{kl},\cr
&\T^{\,i}_{\hskip6pt j} = 
{\delta {\cal L}\over\delta\partial_iA_k} \partial_j A_k -\delta^i_j{\cal L},
\end{align}
we obtain the Hamiltonian and momentum operators
\begin{align}
H_c \equiv \int d^3\theta \, \T^{\,6}_{\hskip8pt6} &= \int d^3\theta\,
\Big( 
- \frac{\sqrt{g}}{e^2} \widetilde G_L^{66}  g^{\alpha\beta}\,F_{6\alpha}
F_{6\beta}
+ \frac{\sqrt{g}}{2e^2} g^{\alpha\alpha'}g^{\beta\beta'} F_{\alpha\beta}
F_{\alpha'\beta'} - \partial_\alpha\Pi^\alpha\, A_6\Big),\label{firstHC}
\end{align}
\begin{align}
P_\alpha\equiv\int d^3\theta \,\T^{6}_{\hskip8pt \alpha} &= 
\int d^3\theta  \Big(- \frac{2}{e^2} \sqrt{g}\,
\widetilde G_{L}^{66} g^{\beta\gamma} F_{6\gamma} F_{\alpha\beta}
-\partial_\beta\Pi^\beta\;A_\alpha + \Pi^6\partial_\alpha A_6 \Big), 
\label{momentum2}\end{align}
where we have integrated by parts; and the conjugate momentum is
\begin{align}
\Pi^\alpha = {\delta\cL\over \delta \partial_6 A_\alpha} = 
-\frac{2}{e^2} \sqrt{g}\, \widetilde G_{L}^{66} g^{\alpha\beta} F_{6\beta} 
-\frac{\theta}{8\pi^2} \epsilon^{\alpha\beta\gamma} F_{\beta\gamma},
\qquad \Pi^6 =  {\delta\cL\over \delta \partial_6 A_6} = 0.
\label{conjmomL} 
\end{align}
Then we have 
\begin{align}
H_c - i \gamma^\alpha P_\alpha = \int d\theta^3 & \Big(
\frac{R_6^2}{4}  \frac{e^2}{\sqrt{g}} g_{\alpha\beta} 
\big(\Pi^\alpha +\frac{\theta}{8\pi^2} \epsilon^{\alpha\gamma\delta} 
F_{\gamma\delta}\big)
\big(\Pi^\beta +\frac{\theta}{8\pi^2} 
\epsilon^{\beta\rho\sigma} F_{\rho\sigma}\big)\cr 
&\hskip6pt + \frac{\sqrt{g}}{2e^2} g^{\alpha\tilde\alpha} g^{\beta\tilde\beta} 
F_{\alpha\beta} F_{\tilde\alpha\tilde\beta} 
-i \gamma^{\alpha} \big(\Pi^\beta + \frac{\theta}{8\pi^2} 
\epsilon^{\beta\gamma\delta} F_{\gamma\delta} \big) F_{\alpha\beta}\Big),
\label{HHamiltonian}\end{align} 
up to terms proportional to $A_6$ and
$\partial_\alpha\Pi^\alpha$ which vanish in Lorenz gauge.
Note the term proportional to $\epsilon^{\beta\gamma\delta}
F_{\gamma\delta} F_{\alpha\beta}$ vanishes
identically.
(\ref{HHamiltonian}) is equal to $H^{4d} -i\gamma^\alpha P^{4d}_\alpha$
given in (\ref{wholepf}), and is used to compute the zero mode
partition function in (\ref{zmpf4d}) via (\ref{maxwell4d}).

To compute the oscillator modes,
the appearance of $\theta$ solely in the combination\hfill\break 
$\Pi^\alpha +\frac{\theta}{8\pi^2} \epsilon^{\alpha\gamma\delta}
F_{\gamma\delta}$ in (\ref{HHamiltonian}) suggests we make
a canonical transformation on the oscillator fields 
$\Pi^\alpha (\vec\theta, \theta^6), A_\beta  (\vec\theta, \theta^6)$
\cite{Kikuchi}. Consider the equal time quantum bracket, suppressing the
$\theta^6$ dependence,

\begin{align}
\left[ \hskip3pt\int d^3\theta' \epsilon^{\alpha\beta\delta} 
F_{\alpha\beta} A_\delta, \hskip5pt\Pi^\gamma({\vec\theta})\hskip3pt \right] 
\hskip2pt= \hskip2pt 2i\epsilon^{\gamma\alpha\beta} 
F_{\alpha\beta}({\vec\theta}), 
\end{align}
and the canonical transformation 
\begin{align}
U(\theta) = {\rm exp} \{ i\frac{\theta}{32\pi^2}\int d^3\theta' 
\epsilon^{\alpha\beta\gamma} F_{\alpha\beta} A_\gamma \},
\label{unitary}\end{align}
under which 
$\Pi^\alpha (\vec\theta, \theta^6), A_\beta  (\vec\theta, \theta^6)$
transform to $\widehat{\Pi}^\alpha (\vec\theta, \theta^6), 
\widehat A_\beta  (\vec\theta, \theta^6)$, 
\begin{align}
\widehat\Pi^\alpha (\vec\theta) = 
U^{-1}({\theta}) \hskip2pt\Pi^\alpha({\vec\theta}) \hskip2pt U(\theta)  
\hskip3pt &= \hskip3pt  \Pi^\alpha ({\vec\theta}) + \frac{\theta}{8\pi^2} 
\epsilon^{\alpha \gamma\delta} F_{\gamma\delta} (\vec\theta)\cr
\widehat A_\beta (\vec\theta) = U^{-1}({\theta}) \hskip2pt A_\beta({\vec\theta}) 
\hskip2pt U(\theta)\hskip3pt &= \hskip3pt  A_{\beta} ({\vec\theta}).
\end{align}
Therefore the exponent (\ref{HHamiltonian}) contains no theta dependence
when written in terms of $\widehat{\Pi}^\alpha$,
which now reads
\begin{align}
\big(H_c - i \gamma^\alpha P_\alpha\big)   
= \int d\theta^3 &\big(-\frac{R_6^2}{4}  \frac{e^2}{\sqrt{g}} g_{\alpha\beta} 
\widehat\Pi^\alpha \widehat\Pi^\beta 
+ \frac{\sqrt{g}}{2e^2} g^{\alpha\tilde\alpha} 
g^{\beta\tilde\beta} F_{\alpha\beta} F_{\tilde\alpha\tilde\beta} 
-i \gamma^{\alpha}  \widehat\Pi^\beta F_{\alpha\beta}\big).
\end{align}
Thus, for the computation of the oscillator partition function
we will quantize with $\theta=0$. Note that had we done this
for the zero modes, it would not be possible to pick the zero mode
integer charges consistently.
Since the zero and oscillator modes commute, we are free to 
canonically transform the latter and not the former.

In the discussion that follows we assume $\theta =0$ and drop the hats. 
We directly quantize the Maxwell theory on the four-torus with the metric 
(\ref{Lfourmetric}) in Lorenz gauge 
using Dirac constraints \cite{Dirac, Das}. 
The theory has a primary constraint $\Pi^6(\vec \theta, \theta^6) \approx 0.$
We can express the Hamiltonian (\ref{firstHC}) in terms of the conjugate
momentum as \begin{align}
H_c  =  \int d\theta^3 &\frac{R_6^2}{4} 
\frac{e^2}{\sqrt{g}} g_{\alpha\beta} \Pi^\alpha \Pi^\beta 
+ \frac{\sqrt{g}}{2e^2} g^{\alpha\tilde\alpha} g^{\beta\tilde\beta} 
F_{\alpha\beta} F_{\tilde\alpha\tilde\beta}.
\label{Hcan}\end{align}
The primary Hamiltonian is defined by
\begin{align}
H_c  =  \int d\theta^3 &\Big( \frac{R_6^2}{4}  \frac{e^2}{\sqrt{g}} 
g_{\alpha\beta} \Pi^\alpha \Pi^\beta + \frac{\sqrt{g}}{2e^2} 
g^{\alpha\tilde\alpha} g^{\beta\tilde\beta} F_{\alpha\beta} 
F_{\tilde\alpha\tilde\beta} - \partial_\alpha\Pi^\alpha A_6 
+ \lambda_1\Pi^6\Big),
\label{Hp}\end{align}
with $\lambda_1$ as a Lagrange multiplier. 
As in \cite{DS}, we use the Dirac method of quantizing with constraints 
for the radiation gauge conditions $A_6\approx 0$,
$\partial^\alpha A_\alpha\approx 0$, and find the equal time commutation 
relations:
\begin{align}
&[\Pi^\beta(\vec\theta, \theta^6), A_\alpha(\vec\theta',\theta^6)] =
- i \Big( \delta^\beta_\alpha - g^{\beta\beta'}
(\partial_\alpha{1\over g^{\gamma\gamma'}\partial_\gamma\partial_{\gamma'}}
\partial_{\beta'})\Big) \;\delta^3 (\theta-\theta'),\cr
& [ A_\alpha(\vec\theta,\theta^6), A_\beta(\vec\theta',\theta^6)]=0,
\qquad [\Pi^\alpha(\vec\theta,\theta^6),\Pi^\beta(\vec\theta',\theta^6)]=0.
\label{thcr}\end{align}
In $A_6=0$ gauge,
the vector potential on the torus is expanded as
\begin{align}
A_\alpha(\vec\theta,\theta^6) = \;{\rm zero\, modes}\; + \sum_{\vec k\ne0, \vec k
\in {\cal Z}_3}
(f_\alpha^{\kappa} a_{\vec k}^{\kappa} e^{ik\cdot \theta}
+ f_\alpha^{\kappa\ast} a_{\vec k}^{\kappa\dagger} e^{-ik\cdot \theta}),
\nonumber\end{align}
where $1\le \kappa\le 2$, $3\le \alpha\le 5$ and $k_6$ defined in (\ref{k6}). The sum is on the dual lattice $\vec k = k_\alpha\in {\cal Z}_3\ne\vec 0.$
Here we only consider the oscillator modes expansion of the potential 
and the conjugate momentum in (\ref{conjmomL}) with vanishing $\theta$ angle
\begin{align}
&A_\alpha(\vec\theta,\theta^6) = \sum_{\vec k\ne 0}
(a_{\vec k\, \alpha} e^{ik\cdot \theta}
+ a_{\vec k\, \alpha}^{\dagger} e^{-ik\cdot \theta}),\cr
&\Pi^\beta(\vec\theta,\theta^6) 
= -i{ 2 \sqrt{g} \over e^2 } \widetilde G_L^{66} g^{\beta\beta'}
\sum_{\vec k} k_6 
\, (a_{\vec k\, \beta'} e^{ik\cdot\theta}
- a_{\vec k\, \beta'}^\dagger e^{-ik\cdot\theta}).
\label{vecfield}\end{align}
and the polarizations absorbed in 
\begin{align}
a_{\vec k\,\alpha}= f_\alpha^\kappa a_{\vec k}^\kappa.\label{polarize}
\end{align}
From (\ref{thcr}), the commutator in terms of the oscillators is 
\begin{align}
\int {d^3\theta d^3\theta'\over (2\pi)^6}
e^{- i k_\alpha\theta^\alpha} e^{- i {k'}_\alpha{\theta'}^\alpha}
[ A_\alpha(\vec\theta, 0), A_\beta(\vec\theta', 0)]
= [(a_{\vec k\, \alpha} + a_{-\vec k\, \alpha}^\dagger),
(a_{\vec k'\, \beta} + a_{-\vec k'\, \beta}^\dagger)] = 0.
\label{Acom}\end{align}
We consider the Fourier transform (\ref{Acom}) of all the commutators 
(\ref{thcr}),
so the commutator of the oscillators is found to be:
\begin{align}
[a_{\vec k\,\alpha}, a^\dagger_{\vec k'\,\beta}] 
&= {e^2 \over 2\sqrt{g}\widetilde G_L^{66} k_6} {1\over 2 (2\pi)^3}
\big(  g_{\alpha\beta} -  {k_\alpha k_\beta\over g^{\gamma\gamma'}k_\gamma k_{\gamma'}} \big)
\delta_{\vec k,\vec k'},\cr
[a_{\vec k\,\alpha}, a_{\vec k'\,\beta}]&=0,\qquad
[a^\dagger_{\vec k\,\alpha}, a^\dagger_{\vec k'\,\beta}]=0.\label{aacom}
\end{align}
In $A_6=0$ gauge, we use (\ref{vecfield}) and (\ref{aacom}) to evaluate the 
Hamiltonian and momentum in (\ref{firstHC}) and (\ref{momentum2})
\begin{align}
H_c &= \int d^3\theta\, {2\sqrt{g}\over e^2}\Big(
-{1\over 2} \widetilde G_L^{66}g^{\alpha\alpha'}\,\partial_6 A_\alpha \partial_6 A_{\alpha'}
+ {1\over 4} g^{\alpha\alpha'}g^{\beta\beta'} F_{\alpha\beta}F_{\alpha'\beta'}\Big),\cr
P_\alpha &=  {2\over R_6^2 e^2}
\int_0^{2\pi} d\theta^3 d\theta^4 d\theta^5
\sqrt{g}\;
g^{\beta\beta'}\,F_{6\beta'}F_{\alpha\beta}.
\label{HPq}\end{align}
With (\ref{vecfield}), (\ref{HPq}) can be expressed in terms of the 
oscillator modes where time-dependent terms cancel,
\begin{align}
H_c &=  (2\pi)^3 {2\sqrt{g}\over e^2}\sum_{\vec k\in {\cal Z}^3 \ne \vec 0}
g^{\alpha\alpha'} |k|^2 \;
(a_{\vec k\,\alpha} a^\dagger_{\vec k\,\alpha'}
+ a^\dagger_{\vec k\,\alpha}a_{\vec k\,\alpha'}),\cr
P_\alpha &= -  {2\sqrt{g}\over e^2}
\widetilde G_L^{66} g^{\beta\beta'} (2\pi)^3
\sum_{\vec k\in{\cal Z}^3 \ne \vec 0}
k_6 k_\alpha \,\big(a_{\vec k\,\beta'} a^\dagger_{\vec k\,\beta}
+ a^\dagger_{\vec k\,\beta'}a_{\vec k\,\beta}\big).
\end{align} 
and we have used the on-shell condition 
$\widetilde G_L^{66} k_6k_6 +|k|^2 =0,$ 
and the transverse condition $k^\alpha a_{\vec k\, \alpha} =
k^\alpha a^\dagger_{\vec k\, \alpha}=0.$ Then,
\begin{align}
- H_c + i\gamma^\alpha P_\alpha & = - i {1\over R_6}  {2 \sqrt{g}\over e^2}
\,(2\pi)^3
\sum_{\vec k\in {\cal Z}^3 \ne \vec 0} |k |\big( - i R_6 |k|
+\gamma^\alpha k_\alpha \big)  g^{\beta\beta'}
\big (a_{\vec k\,\beta} a^\dagger_{\vec k\,\beta'}
+ a^\dagger_{\vec k\,\beta}a_{\vec k\,\beta'}\big).
\end{align}
Inserting the polarizations as $a_{\vec k\,\alpha} = f_\alpha^\kappa a^\kappa_{\vec k}
$ and $a^\dagger_{\vec k\,\alpha} = f_\alpha^{\lambda\ast}
a^{\lambda\dagger}_{\vec k}$ from (\ref{polarize}) in the
commutator (\ref{aacom}) gives
\begin{align}
[a_{\vec k\,\alpha}, a^\dagger_{\vec k'\,\beta}]
&={e^2\over 4\sqrt{g}} {R_6 \over |k|}
{1\over (2\pi)^3} \Big( g_{\alpha\beta} - {k_\alpha k_\beta \over |k|^2}\Big)
\delta_{\vec k,\vec k'}
= f_\alpha^\kappa f_\beta^{\lambda\ast}
[a^\kappa_{\vec k}, a^{\lambda\dagger}_{\vec k}], 
\end{align}
where we choose the normalization
\begin{align}
[a^\kappa_{\vec k}, a^{\lambda\dagger}_{\vec k'}]
= \delta^{\kappa\lambda} \delta_{\vec k,\vec k'},
\label{akappa}\end{align}
with $1 \leq \kappa, \lambda \leq 2$. Then the polarization vectors satisfy

\begin{align}
f_\alpha^\kappa f_\beta^{\lambda\ast} \delta^{\kappa\lambda}
&={e^2 \over 4\sqrt{g}} \frac{R_6}{|k|} 
{1\over (2\pi)^3} \Big( g_{\alpha\beta} 
- {k_\alpha k_\beta \over |k|^2}\Big),\qquad
g^{\beta\beta'}  
f_\beta^\kappa f_{\beta'}^{\lambda\ast} \delta^{\kappa\lambda} =
{e^2 \over 4\sqrt{g}} \frac{R_6}{|k|} 
{1\over (2\pi)^3} \cdot 2,\cr
g^{\beta\beta'}  f_\beta^\kappa f_{\beta'}^{\lambda\ast} &= 
\delta^{\kappa\lambda}
{e^2 \over 4\sqrt{g}} \frac{R_6}{|k|}  {1\over (2\pi)^3}.
\nonumber\end{align}
So the exponent in (\ref{4dpf}) is given by 
\begin{align}
-H_c + i\gamma^\alpha P_\alpha &= - i {R_6} \; {2\sqrt{g}\over e^2}(2\pi)^3
\sum_{\vec k\in{\cal Z}^3 \ne \vec 0} |k| \big( - i R_6 |k|
+ \gamma^\alpha k_\alpha \big)  g^{\beta\beta'}
\big ( 2 a^\dagger_{\vec k\,\beta}a_{\vec k\,\beta'} +
[ a_{\vec k\,\beta},  a^\dagger_{\vec k\,\beta'}] \big)\cr
& = -i \sum_{\vec k\in{\cal Z}^3 \ne \vec 0} \big(
\gamma^\alpha k_\alpha - i R_6 |k|\,\big) \, a_{\vec k}^{\kappa\dagger}
a_{\vec k}^\kappa \quad -{i\over 2}  \sum_{\vec k\in{\cal Z}^3 \ne \vec 0}
\big( -i R_6 |k|\,\big)\,\delta^{\kappa\kappa}.
\label{HplusP}\end{align}
The $U(1)$ partition function is
\begin{align}
Z^{4d, Maxwell} &\equiv tr   \;\exp\{2\pi (- H_c + i\gamma^i P_i)\}
= Z^{4d}_{\rm zero\, modes} \; Z^{4d}_{\rm osc},
\end{align} so from (\ref{HplusP}),
\begin{align}
Z^{4d}_{\rm osc} &
= tr \; e^{ -2\pi i \sum_{\vec k\in {\cal Z}^3 \ne \vec 0} \big(
\gamma^\alpha k_\alpha - i R_6 |k|\,\big) \, a_{\vec k}^{\kappa\dagger}
a_{\vec k}^\kappa  \;\;  - \, \pi R_6 \sum_{\vec k\in{\cal Z}^3 \ne \vec 0}
|k| \,\delta^{\kappa\kappa}}.
\label{ZHplusP}\end{align}
From the usual Fock space argument $$tr\;\omega^{\sum_p p a^\dagger_p a_p}
=\prod_p\sum_{k=0}^\infty \langle k |\omega^{p a^\dagger_p a_p} | k\rangle
=\prod_p {\textstyle 1\over {1 - \omega^p}},$$ 
we perform the trace on the oscillators,
\begin{align}
Z^{4d}_{\rm osc} &=  
\,  \Big( e^{-\pi R_6\sum_{\vec n\in {\cal Z}^3} \sqrt{g^{\alpha\beta}n_\alpha n_\beta}}
\; \; \prod_{\vec n\in {\cal Z}^3\ne \vec 0} {1\over 1 - e^{-i2\pi
(\gamma^\alpha n_\alpha -i R_6\sqrt{g^{\alpha\beta} n_\alpha n_\beta})}}\Big)^2,\label{opf}\\
Z^{4d, Maxwell} &=  Z^{4d}_{\rm zero\;modes} \cdot
\,  \Big( e^{-\pi R_6\sum_{\vec n\in {\cal Z}^3} \sqrt{g^{\alpha\beta}n_\alpha n_\beta}}
\; \; \prod_{\vec n\in {\cal Z}^3\ne \vec 0} {1\over 1 - e^{-2\pi
R_6\sqrt{g^{\alpha\beta} n_\alpha n_\beta} -2\pi i\gamma^\alpha n_\alpha}} \Big)^2,
\label{wpf3}\end{align}
where $Z^{4d}_{\rm zero\;modes}$ is given in (\ref{zmpf4d}).
(\ref{wpf3}) and (\ref{f6doscpf}) are each manifestly 
$SL(3,{\cal Z})$ invariant 
due to the underlying $SO(3)$ invariance we have labeled as $\alpha=3,4,5$.
We use the $SL(3,{\cal Z})$ invariant regularization of the
vacuum energy reviewed in Appendix C to obtain 
\begin{align}
Z^{4d, Maxwell}&=  Z^{4d}_{\rm zero\;modes} \cdot
\,  \Big( e^{{1 \over 2}R_6\pi^{-2}
\sum_{\vec n\ne 0} {\sqrt{\tilde g}\over 
({g_{\alpha\beta}n^\alpha n^\beta})^{2}}}
\; \; \prod_{\vec n\in {\cal Z}^3\ne \vec 0} {1\over 1 - e^{-2\pi
R_6\sqrt{g^{\alpha\beta} n_\alpha n_\beta} - 2\pi i \gamma^\alpha n_\alpha}} 
\Big)^2,
\label{wpf4}\end{align} which leads to (\ref{f4f}).

On the other hand, one can evaluate the oscillator trace for the $6d$ 
chiral boson from (\ref{6dpf}) as in\cite{DN},\cite{DS}. 
The exponent in the trace is
\begin{align} -2\pi R_6{\cal H} +i2\pi\gamma^iP_i&= 
{i\pi\over 12}\int_0^{2\pi} d^5\theta H_{lrs}
\epsilon^{lrsmn} H_{6mn}
= {i\pi\over 2}\int_0^{2\pi} d^5\theta \sqrt{-G}
H^{6mn} H_{6mn}\cr
&={-i\pi \int_0^{2\pi} d^5\theta
(\Pi^{mn} H_{6mn} + H_{6mn} \Pi^{mn})}\cr
&=-2 i \pi\sum_{\vec p \ne 0} p_6
{\cal C}_{\vec p}^{\kappa\dagger} B_{\vec p}^\kappa
-i \pi \sum_{\vec p \ne 0} p_6 \delta^{\kappa\kappa},
\label{chone}\end{align}
where $\Pi^{mn}= -{\sqrt{-G}\over 4} \Pi^{6mn}$, and $\Pi^{6mn}$ 
is the momentum conjugate to $B_{MN}$. In the gauge $B_{6n}=0,$
the normal mode expansion for the free quantum fields $B_{mn}$ and $\Pi^{mn}$
on a torus is given in terms of oscillators $B_{\vec p}^\kappa$ and
${\cal C}_{\vec p}^{\kappa\dagger}$ 
defined in \cite{DN}, with the commutation relations
\begin{align}
[B_{\vec p}^\kappa, {\cal C}_{\vec p'}^{\lambda\dagger}]= 
\delta^{\kappa\lambda} \; \delta_{\vec p,\vec p'}
\label{BCcr}\end{align}
where $1 \leq \kappa, \lambda \leq 3$ labels the three physical degrees of
freedom of the chiral two-form, and ${\vec p}=(p_1, p_2, p_{\alpha})$
lies on the integer lattice ${\cal Z}^5$. 
From the on-shell condition $G^{LM} p_Lp_M=0$, 
\begin{align}
p_6 = -\gamma^{\alpha} p_\alpha - i R_6 \sqrt{ g^{\alpha\beta} p_\alpha p_\beta + \frac{p_1^2}{R_1^2} + (\frac{1}{R_2^2} + \frac{{\beta^2}^2}{R_1^2} ) p_2^2 + 2\frac{\beta^2}{R_1^2} p_1 p_2 }  .\end{align}
Thus the 
oscillator partition function of the chiral two-form on $T^2 \times T^4$ 
is obtained by tracing over the oscillators
\begin{align}
Z^{6d}_{\rm osc} &
= tr \; e^{-2 i \pi\sum_{\vec p \ne 0} p_6
{\cal C}_{\vec p}^{\kappa\dagger} B_{\vec p}^\kappa
-i \pi \sum_{\vec p \ne 0} p_6 \delta^{\kappa\kappa}}\cr
&= \Big(e^{-\pi R_6 \sum_{\vec p} \sqrt{g^{\alpha\beta} p_{\alpha}p_{\beta} 
+ \tilde p^2}}
\prod_{\vec p\neq 0} \frac{1}{1-e^{-2\pi ip_6}}\Big)^3\cr
&=  \Big(e^{-\pi R_6 \sum_{\vec p\in {\cal Z}^5} 
\sqrt{g^{\alpha\beta} p_{\alpha}p_{\beta} 
+ \tilde p^2}}
\prod_{\vec p\in {\cal Z}^5\neq \vec 0} {\textstyle 1\over
{1- e^{-2\pi R_6 \sqrt{g^{\alpha\beta} p_\alpha p_\beta +\tilde p^2} 
+ 2\pi i\gamma^\alpha p_\alpha}}}\Big)^3,
\label{ZHplusP6d}\end{align}
where $\tilde p^2 \equiv \frac{p_1^2}{R_1^2} +
(\frac{1}{R_2^2} + 
\frac{{\beta^2}^2}{R_1^2} ) p_2^2 + 2\frac{\beta^2}{R_1^2} p_1 p_2 $.
Regularizing the vacuum energy in the oscillator sum \cite{DN}
yields
{\begin{align}
Z^{6d, chiral} &= Z^{6d}_{\rm zero\,modes} \cdot
\Bigl ( e^{ R_6 \pi^{-3} \sum_{\vec n\ne \vec 0} {\sqrt{G_5}\over
(G_{mp} n^mn^p)^{\,3}}} \prod_{\vec p\in {\cal Z}^5\ne \vec 0}
{\textstyle 1\over{1- e^{-2\pi R_6 \sqrt{g^{\alpha\beta} p_\alpha p_\beta 
+\tilde p^2} + 2\pi i\gamma^\alpha p_\alpha}}}\Bigr )^3,
\label{f6doscpf}\end{align}
\vskip-10pt
where $\vec n\in {\cal Z}^5$ is on the dual lattice, $G_{mp}$ is defined in
(\ref{metric1}), 
and $Z^{6d}_{\rm zero\,modes}$ is given in (\ref{zmpf}). 

Comparing the $4d$ and $6d$ oscillator traces (\ref{wpf3}) and  
(\ref {ZHplusP6d}), the $6d$ chiral boson sum
has a cube rather than a square, corresponding to one additional polarization, 
and it contains Kaluza-Klein modes. 
In Appendix D, we prove that the product of the zero mode 
and the oscillator mode partition function for the $4d$ theory in 
(\ref{wpf4}) is $SL(4,{\cal Z})$ invariant.
In (\ref{wpf4again}) we give an equivalent expression,
\begin{align}
&Z^{4d, Maxwell}=  Z^{4d}_{\rm zero\;modes} \cdot
\Big( e^{\pi R_6\over 6 R_3} \prod_{n_3 \ne 0} {1\over
1 - e^{-2\pi  {R_6\over R_3}|n_3| +2\pi i \gamma^3 n_3}}\Big )^2
\cr& \hskip1pt \cdot
\Big( \prod_{(n_a)\in {\cal Z}^2\ne (0,0)} e^{-2\pi R_6 <H>_{p_\perp}}
\; \; \prod_{n_3\in {\cal Z}} {1\over 1 - e^{-2\pi
R_6\sqrt{g^{\alpha\beta} n_\alpha n_\beta} + 2\pi i\gamma^\alpha n_\alpha}} \Big)^2,\cr
\label{wpfyetagain}\end{align}
where $ 4\leq a \leq 5$, with $ <H>_{p_\perp}$ defined in (\ref{C3}).

In Appendix D, we also prove the $SL(4, {\cal Z})$ invariance of 
the $6d$ chiral partition function 
(\ref{f6doscpf}), using the equivalent form (\ref{a6dpf}),
\begin{align}
&Z^{6d, chiral} = Z^{6d}_{\rm zero\,modes}\cdot\bigl (e^{\pi R_6\over 6 R_3}
\prod_{n_3 \in {\cal Z} \ne 0} {1\over
1 - e^{- 2\pi {R_6\over R_3} |n_3| + 2\pi i \,\gamma^3 \,n_3\,)}}\bigr )^3\cr
&\hskip2pt \cdot
\bigl(\prod_{n_\perp \in {\cal Z}^4\ne (0,0,0,0)}
e^{-2\pi R_6 < H>^{6d}_{p_\perp}} \prod_{n_3\in {\cal Z}}
{1\over 1-e^{-2\pi R_6\sqrt{g^{\alpha\beta} n_\alpha n_\beta + {\tilde n}^2}
\; +i2\pi\gamma^\alpha n_\alpha}}\bigr)^3,\label{wpfmore}\end{align}
with $< H>^{6d}_{p_\perp}$ in (\ref{BES6}), and
${\tilde n}^2 = {(n_1)^2\over R_1^2}  + (\frac{1}{R_2^2} +\frac{(\beta^2)^2}
{R_1^2})n_2^2   + 2\frac{\beta^2}{R_1^2} n_2 n_1.$
In the limit when $R_1$ and $R_2$ are 
small with respect to the metric parameters $g_{\alpha\beta}, R_6$ 
of the four-torus, 
the contribution from each polarization in 
(\ref{wpfyetagain}) and (\ref{wpfmore}) is equivalent. 
To see this limit, we can
separate the product on $ n_\perp = (n_1, n_2, n_a)\ne 0_\perp $ 
in (\ref{wpfmore}),
into 
$(n_1=0, n_2 = 0, n_a\ne (0,0))$, $(n_1 \ne 0, n_2 \ne 0, \hskip2pt\hbox{all}\; n_a)$, $(n_1=0, n_2\ne 0,\hskip2pt\hbox{all}\; n_a )$,  $(n_1 \neq 0, n_2 =0,  \hskip2pt\hbox{all}\; n_a ))$
to find, at fixed $n_3$,
\begin{align}
&\prod_{n_\perp\in{\cal Z}^4\ne (0,0,0,0)}
{\textstyle 1\over{1- e^{-2\pi R_6 \sqrt{g^{\alpha\beta} n_\alpha n_\beta + 
{(n_1)^2\over R_1^2}  + (\frac{1}{R_2^2} +\frac{(\beta^2)^2}{R_1^2})n_2^2   
+ 2\frac{\beta^2}{R_1^2} n_2 n_1 }
+ 2\pi i\gamma^\alpha n_\alpha}}}\cr
&\hskip40pt= \prod_{n_a\in {\cal Z}^2\ne (0,0)}
{\textstyle 1\over{1- e^{-2\pi R_6 \sqrt{g^{\alpha\beta} n_\alpha n_\beta } + 2\pi i \gamma^\alpha n_\alpha
}}}\cr &\hskip40pt\cdot \prod_{n_1\ne 0, n_2 \neq 0, (n_a\in {\cal Z}^2)}
{\textstyle 1\over{1- e^{-2\pi R_6 \sqrt{g^{\alpha\beta} n_\alpha n_\beta 
+ {(n_1)^2\over R_1^2} + (\frac{1}{R_2^2} +\frac{(\beta^2)^2}{R_1^2})n_2^2   
+ \frac{\beta^2}{R_1^2} n_2 n_1 }
+ 2\pi i\gamma^\alpha n_\alpha}}}\cr
&\hskip40pt \cdot \prod_{n_1= 0, n_2 \neq 0, (n_a\in {\cal Z}^2)}
{\textstyle 1\over{1- e^{-2\pi R_6 \sqrt{g^{\alpha\beta} n_\alpha n_\beta  
+ (\frac{1}{R_2^2} +\frac{(\beta^2)^2}{R_1^2})n_2^2}
+ 2\pi i\gamma^\alpha n_\alpha}}}\cr
\cr &\hskip40pt \cdot \prod_{n_2 = 0, n_1 \neq 0, (n_a\in {\cal Z}^2)}
{\textstyle 1\over{1- e^{-2\pi R_6 \sqrt{g^{\alpha\beta} n_\alpha n_\beta + {(n_1)^2\over R_1^2} }
+ 2\pi i\gamma^\alpha n_\alpha}}}\cr
\end{align}
Thus for $T^2$ smaller than $T^4$, the last three products reduce to unity, so
\begin{align}
&\prod_{n_\perp\in{\cal Z}^4\ne {\vec 0}}
{\textstyle 1\over{1- e^{-2\pi R_6 \sqrt{g^{\alpha\beta} n_\alpha n_\beta 
+ {\tilde n}^2}
+2\pi i\gamma^\alpha n_\beta}}}\xrightarrow {R_1,R_2 \rightarrow 0} 
\prod_{n_a\in {\cal Z}^2\ne (0,0)}
{\textstyle 1\over{1- e^{-2\pi R_6\sqrt{g^{\alpha\beta} n_\alpha n_\beta } 
+ 2\pi i\gamma^\alpha n_\alpha
}}}.
\label{65lim}\end{align} 
The regularized vacuum energies in (\ref{C3}) and (\ref{BES6}),
\begin{align}
<H>_{p_\perp\ne 0}&=-{\pi}^{-1}\; |p_\perp|
\sum_{n=1}^\infty\cos(p_a\kappa^a 2\pi n)
{K_1(2\pi n R_3 |p_\perp|)\over n}, \quad
\hbox{for}\quad
|p_\perp|\equiv \sqrt{\widetilde g^{a b} n_a n_b},\cr
<H>^{6d}_{p_\perp\ne 0}&=-{\pi}^{-1}\; |p_\perp|
\sum_{n=1}^\infty\cos(p_a\kappa^a 2\pi n)
{K_1(2\pi n R_3 |p_\perp|)\over n}, \quad\hbox{for}\quad|p_\perp|
\equiv \sqrt{   {\tilde n}^2+ \widetilde g^{ab} n_a n_b}, \end{align}
have the same form of spherical Bessel function, 
but the argument differs by modes $(p_1, p_2)$.  
Again separating the product on $n_\perp = (n_1, n_2, n_a)$ in (\ref{wpfmore}),
into \hfill\break
$(n_1=0, n_2 = 0, n_a\ne (0,0))$, $(n_1 \ne 0, n_2 \ne 0 \hskip2pt\hbox{all}\; n_a)$, $(n_1=0, n_2\ne 0,\hskip2pt\hbox{all}\; n_a )$,  $(n_1 \neq 0, n_2 =0,  \hskip2pt\hbox{all}\; n_a ))$
we have
\begin{align}
&\prod_{n_\perp\in {\cal Z}^4\ne (0,0,0,0)}
e^{-2\pi R_6 < H>^{6d}_{p_\perp}} =
\bigl(\prod_{n_a\in {\cal Z}^2\ne (0,0)}
e^{-2\pi R_6 < H>^{4d}_{p_\perp}}\bigr)\cdot
\bigl(\prod_{n_1\ne 0, n_2\ne 0, n_a\in{\cal Z}^2}
e^{-2\pi R_6 < H>^{6d}_{p_\perp}}\bigr)\cr
&\hskip25pt\cdot\bigl(\prod_{n_1\ne 0, n_2 = 0, n_a\in{\cal Z}^2}
e^{-2\pi R_6 < H>^{6d}_{p_\perp}}\bigr) \cdot\bigl(\prod_{n_1=0, n_2\ne 0, n_a\in{\cal Z}^2}
e^{-2\pi R_6 < H>^{6d}_{p_\perp}}\bigr)\label{relH}\end{align}
In the limit $R_1, R_2\rightarrow 0$, the last three products are unity. For example, the second is unity because
for $n_1, n_2\ne 0$, \begin{align}
&\lim_{R_1, R_2\rightarrow 0}  \sqrt{{\tilde n}^2 
+ \widetilde g^{\alpha\beta} n_\alpha n_\beta} \sim \sqrt{  {\tilde n}^2 },\cr 
&\lim_{R_1, R_2\rightarrow 0}(  |p_\perp| \; K_1(2\pi n R_3 |p_\perp|) 
= \lim_{R_1, R_2\rightarrow 0}  \sqrt{{\tilde n}^2}\: K_1
\Big (2\pi n R_3    \big(\sqrt{ {\tilde n}^2 }      \big)\Big)=0,
\end{align} 
since $\lim_{x\rightarrow\infty} x K_1(x) \sim \sqrt{x} \;e^{-x} \rightarrow 0$
\textcolor{blue}{\cite{AS}}. So (\ref{relH}) leads to 
\begin{align}
\lim_{R_1, R_2\rightarrow 0} \prod_{n_\perp\in{\cal Z}^4\ne (0,0,0,0)}
e^{-2\pi R_6 < H>^{6d}_{p_\perp}} = 
\prod_{n_a\in {\cal Z}^2\ne (0,0)} e^{-2\pi R_6 < H>_{p_\perp}}.
\end{align} 
Thus in the limit when $T^2$ is
small with respect to $T^4$, 
\begin{align}
\lim_{R_1, R_2\rightarrow 0} & \prod_{n_\perp\in{\cal Z}^4\ne (0,0,0,0)}
e^{-2\pi R_6 < H>^{6d}_{p_\perp}} \prod_{n_3\in {\cal Z}}
{1\over 1-e^{-2\pi R_6\sqrt{(g^{\alpha\beta} n_\alpha n_\beta 
+ {n^2_1\over R_1^2}  
+({1\over R_2^2} + \frac{(\beta^2)^2}{R_1^2})n_2^2   
+ 2\frac{\beta^2}{R_1^2} n_2 n_1}   
\; +i2\pi\gamma^\alpha n_\alpha}}\cr&
= \prod_{n_a\in {\cal Z}^2\ne (0,0)} e^{-2\pi R_6 <H>_{p_\perp}}
\; \; \prod_{n_3\in {\cal Z}} {1\over 1 - e^{-2\pi
R_6\sqrt{g^{\alpha\beta} n_\alpha n_\beta} + 2\pi i\gamma^\alpha n_\alpha}}.
\end{align}
So we have shown the partition functions
of the chiral theory on $T^2\times T^4$ and of gauge theory on $T^4$,
agree in the small $T^2$ limit upon neglecting the less interesting 
contribution $\epsilon'$,
\begin{align}
\lim_{R_1, R_2\rightarrow 0} \; Z_{osc}^{6d} 
= \epsilon' \cdot Z_{osc}^{4d},
\end{align}
which is (\ref{osclim}).
Again, $\epsilon'$ is equivalently the oscillator 
contribution from one polarization, that is
\begin{align}
\epsilon' =  \Big( e^{{1 \over 8}R_6\pi^{-2}
\sum_{\vec n\ne 0} {\sqrt{\tilde g}\over 
({g_{\alpha\beta}n^\alpha n^\beta})^{2}}}
\; \cdot \prod_{\vec n\in {\cal Z}^3\ne \vec 0} {1\over 1 - e^{-2\pi
R_6\sqrt{g^{\alpha\beta} n_\alpha n_\beta} - 2\pi i \gamma^\alpha n_\alpha}} 
\Big)\end{align}

The relation between the $4d$ gauge and $6d$ tensor partition function
is shown in the small $T^2$ limit,
\begin{align}
\lim_{R_1, R_2\rightarrow 0} \; Z^{6d, chiral} = 
\epsilon \epsilon' \cdot Z^{4d, Maxwell},
\end{align}
which is (\ref{llim4d}).
$\epsilon\epsilon'$ is the partition function of a real scalar field in
4d, and is independent of the gauge coupling $\tau.$ 

\section{\bf S-duality of $Z^{4d,Maxwell}$ from $Z^{6d,chiral}$}

In Appendices B and D  we show explicitly
how the $SL(2,{\cal Z})\times SL(4,{\cal Z})$
symmetry of the partition function of the $6d$ tensor field
of the M-fivebrane of $N=(2,0)$ theory compactified on $T^2\times T^4$
implies the $SL(2,{\cal Z})$ S-duality of the $4d$ $U(1)$ gauge field
partition function. These computations use the Hamiltonian formulation.
In Appendix A we review the path integral formalism for the $4d$ 
zero and non-zero mode partition functions, 
and give their relations to the quantities computed in the Hamiltonian
formulation. 
The results are summarized here.
\begin{align}
Z_{\rm zero \; modes}^{4d} 
&= ({\rm Im}\,\tau)^{\frac{3}{2}} {{g}^{\frac{1}{4}}\over 
R_6^2} \, Z^{PI}_{{\rm zero\, modes}}.
\label{zmzb}\end{align}
\begin{align}
Z^{4d}_{osc} = ({\rm Im}\,\tau)^{-{\frac{3}{2}}}
g^{-{1\over 4}} R_6^2 Z^{PI}_{osc}.  
\label{relationosc}\end{align}
\begin{align}
&Z^{4d}_{\rm zero\; modes}\longrightarrow Z^{4d}_{\rm zero \; modes},\qquad 
Z^{PI}_{\rm zero\; modes} \longrightarrow |\tau|^{3}
Z^{PI}_{\rm zero \; modes} 
\hskip35pt \hbox{under $S$}\cr
&Z^{4d}_{\rm zero\; modes} \longrightarrow Z^{4d}_{\rm zero \; modes},
\qquad
Z^{PI}_{\rm zero\; modes} \longrightarrow Z^{PI}_{\rm zero \; modes}
\qquad\hskip23pt
\hskip8pt\hbox{under $T$}
\label{agzmtransf}\end{align}
and
\begin{align}
& Z^{4d}_{\rm osc} \longrightarrow Z^{4d}_{\rm osc},\qquad
Z^{PI}_{\rm  non-zero \; modes} \longrightarrow |\tau|^{-3} 
Z^{PI}_{\rm non-zero \; modes} 
\hskip35pt \hbox{under $S$}\cr
& Z^{4d}_{\rm osc}\longrightarrow Z^{4d}_{\rm osc}, \hskip23pt
Z^{PI}_{\rm  non-zero \; modes} \longrightarrow 
Z^{PI}_{\rm  non-zero \; modes}
\hskip58pt \hbox{under $T$}.
\label{agosctransf}\end{align}
$S$ and $T$ are the generators of the duality symmetry $SL(2,{\cal Z})$,
$S:\tau\rightarrow -{1\over \tau}$, $T:\tau\rightarrow\tau-1$, where
$\tau = {\theta\over 2\pi} + i{4\pi\over e^2}$ is also given by 
the modulus of the two-torus, $\tau = \beta^2+i{R_1\over R_2}.$

\section{\bf Conclusions and Discussion}

We computed the partition function of the abelian gauge theory on a general 
four-dimensional torus $T^4$ and the partition function of a chiral boson  
compactified on $T^2\times T^4$. The coupling for the 
$4d$ gauge theory, $\tau =\frac{\theta}{2\pi} +  i\frac{4\pi}{e^2}$, 
is identified with the complex modulus $\tau = \beta^2 + i\frac{R_1}{R_2}$
of the two-torus $T^2$ in directions 1 and 2. 
Assuming the metric of $T^2$ is much smaller than $T^4$, 
the $6d$ partition function factorizes to a partition function for 
gauge theory on $T^4$ and a contribution from the extra scalar
arising from compactification.\footnote{The Lagrangian for this single $4d$ 
scalar with a Lorentzian signature metric is \hfill\break
${\cal L}= {R_6\sqrt{\tilde g}\over R_1R_2}
\, ({1\over 2 R^2_6}\partial_6\phi\partial_6\phi - {1\over 2}
g^{\alpha\beta} \partial_\alpha\phi\partial_\beta\phi)$.} 
The $6d$ partition function has a manifest $SL(2, {\cal Z}) \times 
SL(4, {\cal Z})$ symmetry. Therefore 
the $SL(2,{\cal Z})$ symmetry with the group action on the coupling,
$\tau = \frac{\theta}{2\pi} + i\frac{4\pi}{e^2}$,  
known as S-duality becomes manifest in the $4d$ Maxwell theory.
Presumably this happens for an arbitrary four manifold, but we chose
$T^4$ in order to generate explicit formulas, {\it i.e.} explicit
functions of $\tau$ and the 4d metric.

The $6d$ chiral two-form has no Lagrangian, 
so we use the Hamiltonian approach to compute
both the $4d$ and $6d$ partition 
functions. For $4d$ gauge theory, the integration of the electric and magnetic 
fields as observables around one- and two-cycles respectively take integer 
values due to charge quantization. We sum over all possible 
integers to get the zero mode partition function. For the oscillator modes, 
we quantize the gauge theory using the Dirac method with 
constraints. In $6d$, the partition function follows from \cite{DN},\cite{DS}.

We have also given the path integral result for the $4d$ partition function.
It agrees with the partition function obtained in 
the Hamiltonian formulation. However, the path integral 
factors into zero modes and oscillator modes differently, 
which leads to different $SL(2,{\cal Z})$ transformation properties for
the components. 
The $6d$ and $4d$ partition functions share the same 
$SL(2,{\cal Z})\times SL(4,{\cal Z})$ symmetry.


If we consider supersymmetry,
compactification of the $6d$ theory on $T^2$ leads to $N=4$ gauge theory
in the limit of small $T^2$.
On the other hand, an $N=2$ theory of class 
{\cal S} \cite{Gaiotto},\cite{Neitzke} arises when the $6d$, $(2,0)$ theory
is compactified on a punctured Riemann surface with genus $g$.  
Here the  mapping class group of the
Riemann surfaces acts as a generalized S-duality on 4d super-Yang-Mills 
theory \cite{AGT}-\cite{Tachikawa}. 

In 
another direction, we can study the $2d$ field theory
present when $6d$ theory is compactified on a four-dimensional manifold. 
A $2d$-$4d$ correspondence, relating a generalized gauge theory partition
function and a 2d correlation function, acting 
between $N=4$ gauge theory on $S^4$ and a
$2d$ Toda-Liouville conformal theory on $T^2$, holds for the radius of
$S^4$ fixed to 1 \cite{AGT},\cite{Pestun},\cite{Okuda}-\cite{NO}.
It is difficult to see how the $2d$-$4d$ correspondence works for 
the gauge field on $T^4$ because the $4d$ oscillator partition function
is most naturally viewed as that of a $2d$ theory on a $T^2$ in directions
3 and 6, whereas the $4d$ zero mode sum is equivalent to a $2d$ zero mode
partition function on a $T^2$ in directions 1 and 2. For an arbitrary 
$4d$ metric, the theory may be too rich for a $2d$-$4d$ pairing.
A $2d$-$4d$ relation can also be analyzed 
from a topological point of view \cite{VafaWitten},\cite{Gukov1},\cite{Gukov2}. 
Finding explicit results, such as we have derived for $T^2\times T^4$,
for these more general investigations would be advantageous. 
\vfill\eject

\appendix
\section{\bf Comparison of the $4d$ $U(1)$ partition function in the 
Hamiltonian and path integral formulations}

For convenience in comparing the $4d$ gauge theory with the $6d$ chiral 
theory in sections 2 and 3, we quantized both using canonical 
quantization. 
Since a Lagrangian exists for the $4d$ gauge theory, it is useful to 
verify that its path integral quantization agrees with canonical quantization.
We find the two quantizations 
distribute zero and oscillator mode contributions differently,
and thus these factors transform differently under the action of 
$SL(2,{\cal Z})$. 
We summarize the path integral quantization results from 
\cite{Verlinde}, \cite{ Zucchini}, \cite{Witten0}, \cite{Etesi}.
Following \cite{Verlinde}, \cite{Witten0}, the two-form zero mode part, 
$F \over 2\pi$ is the harmonic 
representative and can be expanded in terms of the basis $\alpha_I 
= {1\over (2\pi)^2}
d\theta^1 \wedge d\theta^2$, etc., $I = 1,2,..,6$ namely
\begin{align}
\frac{F}{2\pi} \equiv m = \sum_{I} m_I \alpha_I,
\label{ints}\end{align}
where $m_I$ are integers.
Define $(m,n)$ to be the intersection form such that $(m,n)=\int m \wedge n$,
and thus
\begin{align}
(m, m) &= \frac{1}{16\pi^2} \int d^4 \theta \epsilon^{ijkl} F_{ij} F_{kl}\cr
(m, *m) &= \frac{1}{8\pi^2} \int d^4  \theta \sqrt{g} F^{ij} F_{ij}.
\end{align}So the action (\ref{actiontheta2}) is given as
\begin{align}
I &= \frac{4\pi^2}{e^2} (m, *m) - \frac{i\theta}{2} (m,m) = \frac{1}{2e^2} 
\int d^4\theta \sqrt{g} F^{ij} F_{ij} - \frac{i\theta}{32\pi^2} 
\int d^4\theta \epsilon^{ijkl} F_{ij} F_{kl}.
\label{actionm}\end{align}
The zero mode partition function from the path integral formalism can be 
expressed as a lattice sum over the integral basis of $m_I$ \cite{Verlinde},
\cite{Witten0},
\begin{align}
Z_{\rm zero \; modes} ^{PI} &= \sum_{m_I \in {\cal Z}^6} {\rm exp} 
\Big[ -\frac{4\pi^2}{e^2} \big(m, *m \big) + \frac{i\theta}{2} \big(m,m \big)  
\Big] \cr &=  \sum_{m_I \in {\cal Z}^6} {\rm exp} \Big[ \frac{i \pi}{2} \tau 
\Big(\big(m,m\big) + \big(m,*m\big) \Big) - \frac{i\pi}{2} {\bar \tau} 
\Big(-\big(m,m\big)+\big(m,*m\big)\Big) \Big],\;\end{align}
where $\tau = {\theta\over 2\pi} + i {4\pi\over e^2}$, and we have
chosen the $\theta$ dependence of the action as in \cite{Verlinde}.
Alternatively the zero mode sum be can written in terms of the 
metric using (\ref{actionm})
\begin{align}
{Z}_{\rm zero \; modes}^{PI} &=
\sum_{\widetilde F_{ij} 
\in {\cal Z}^6} {\rm exp} \big\{ \big [- \frac{\pi}{2} R_6 \sqrt{\tilde g} 
g^{\alpha\beta} g^{\gamma\delta} 
\widetilde F_{\alpha\gamma} {\tilde F}_{\beta\delta}
-\pi  \frac{\sqrt{\tilde g}}{R_6} g^{\delta\delta'} 
{\widetilde F}_{\delta\beta} \gamma^{\beta} {\widetilde F}_{\delta'\beta'} 
\gamma^{\beta'} -\pi\frac{\sqrt{\tilde g}}{R_6} g^{\alpha\beta} 
{\widetilde F}_{6\alpha} {\widetilde F}_{6\beta} \cr 
&\hskip90pt + 2\pi \frac{\sqrt{\tilde g}}{R_6} 
g^{\alpha\delta} {\widetilde F}_{6\alpha} {\widetilde F}_{\delta\beta} 
\gamma^{\beta} - i {\theta e^2\over 8\pi} \epsilon^{\alpha\beta\gamma} 
{\widetilde F}_{6\alpha} {\tilde F}_{\beta\gamma} \big] \,{4\pi\over e^2}\big\}
\label{f2}\end{align}
where ${\widetilde F}_{ij} = 2\pi F_{ij} = m_I$ are integers due to 
the charge quantization (\ref{ints}), and where we have taken into account
the integrations $\int d^4\theta = (2\pi)^4$ in (\ref{f2}).
To compare the zero mode partition functions from 
the Hamiltonian and path integral formalisms, 
we rewrite the Hamiltonian formulation result (\ref{zmpf4d}) as
\begin{align}
&Z^{4d}_{\rm zero \; modes}\cr
&=\sum_{{\widetilde \Pi}^\alpha, {\widetilde F}_{\alpha\beta}} {\rm exp}
\Big[ -\frac{e^2 R_6}{4\sqrt{\tilde g}} g_{\alpha\beta} 
\big(\widetilde \Pi^{\alpha} + i \frac{4\pi\sqrt{\tilde g}}{e^2R_6} 
g^{\alpha\delta} {\widetilde F}_{\delta\lambda}\gamma^{\lambda} 
+ \frac{\theta\epsilon^{\alpha\gamma\delta}}{4\pi} 
{\widetilde F}_{\gamma\delta}
\big)\cdot\big(\widetilde \Pi^{\beta} + i 
\frac{4\pi\sqrt{\tilde g}}{e^2R_6} g^{\beta\delta'} 
\widetilde F_{\delta'\lambda'}\gamma^{\lambda'}  
+ \frac{\theta\epsilon^{\beta\gamma'\delta'}}{4\pi} 
{\widetilde F}_{\gamma'\delta'} \big) \cr 
&\hskip70pt -{4\pi^2\over e^2} \frac{\sqrt{\tilde g}}{R_6} 
g^{\delta\delta'} {\widetilde F}_{\delta\beta}\gamma^{\beta} 
\widetilde F_{\delta'\beta'}\gamma^{\beta'} - 
{2\pi^2\over e^2}\sqrt{g}  g^{\alpha\beta} g^{\gamma\delta} 
{\widetilde F}_{\alpha\gamma} {\widetilde F}_{\beta\delta} \Big].\end{align}
After Poisson resummation,  \begin{align}
\sum_{n\in {\cal Z}^3} {\rm exp} [-\pi (n+ x) \cdot {\cal A} \cdot (n + x)] 
= (\rm det {\cal A})^{-\frac{1}{2}} \sum_{n\in {\cal Z}^3} 
e^{-\pi n\cdot {\cal A}^{-1} \cdot n} e^{2\pi i n\cdot x},
\end{align}
where ${\cal A}_{\alpha\beta}  \equiv {e^2 R_6\over 4\pi \sqrt{\tilde g}} 
g_{\alpha\beta}$ and $x^{\alpha} \equiv i 
\frac{4\pi \sqrt{\tilde g}}{e^2 R_6} g^{\alpha\delta} 
{\widetilde F}_{\delta\lambda} \gamma^\lambda  
+ \frac{\theta}{4\pi} 
\epsilon^{\alpha\gamma\delta} \widetilde F_{\gamma\delta}$,  
we get the Hamiltonian expression as
\begin{align}
Z_{\rm zero \; modes}^{4d}&= ({e^2\over 4\pi})^{-\frac{3}{2}} 
\frac{{\widetilde g}^{\frac{1}{4}}}{ {R_6}^{\frac{3}{2}}}
\sum_{\hat\Pi_\alpha,\tilde F_{\alpha\beta}} {\rm exp} 
\{-\frac{4\pi^2\sqrt{\tilde g}}{e^2 R_6} g^{\alpha\beta} {\hat \Pi}_{\alpha} 
{\hat \Pi}_{\beta} - i{\theta\over 2} \hat\Pi_\alpha 
\epsilon^{\alpha\gamma\delta} \tilde F_{\gamma\delta} 
+ \frac{8\pi^2\sqrt{\tilde g}}{e^2R_6} g^{\alpha\beta} 
\hat\Pi_\alpha \tilde F_{\beta\delta} \gamma^{\delta} \cr
&\hskip110pt
-\frac{4\pi^2}{e^2} \frac{\sqrt{\tilde g}}{R_6} g^{\delta\delta'} 
\tilde F_{\delta\beta}\gamma^{\beta} \tilde F_{\delta'\beta'}\gamma^{\beta'} 
- \frac{2\pi^2R_6}{e^2} \sqrt{\tilde g}  g^{\alpha\beta} 
g^{\gamma\delta} {\tilde F}_{\alpha\gamma} {\tilde F}_{\beta\delta} \}\cr
&=({\rm Im}\,\tau)^{\frac{3}{2}} \frac{{\widetilde g}^{\frac{1}{4}}}
{ {R_6}^{\frac{3}{2}}} Z^{PI}_{{\rm zero\, modes}},
\label{zmzm}\end{align}
where $\hat \Pi_\alpha$ is the integer value of $\widetilde \Pi^\alpha$,
and we identify  
$\hat \Pi_\alpha$ with ${\widetilde F}_{6\alpha}$ in (\ref{f2}).
Then \begin{align}
{Z}_{\rm zero \; modes}^{PI}
= ({\rm Im}\,\tau)^{-\frac{3}{2}}\frac{R_6^2}
{g^{\frac{1}{4}}}\,{Z}_{\rm zero\; modes}^{4d},
\label{relationzero}\end{align}
which is (\ref{zmzb}).

We review from \cite{Zucchini} how the non-zero mode partition function is 
defined by a path integral, 
\begin{align}
Z^{PI}_{\rm non-zero\, modes} = \int_A  \hskip5pt DA^\mu 
e^{-I}.
\end{align}
Performing the functional integration with the Fadeev-Popov approach, 
\cite{Zucchini} regularizes the path integral by 
\begin{align}
Z^{PI}_{\rm non-zero\, modes}  = 
\frac{1}{(2\pi)^{\frac{b_1 -1}{2}}} \left(\frac{g}
{{\rm vol} T^4}\right)^{\frac{1}{2}}
\Big[ {\rm det}(\Delta_0) \frac{{\rm det}( 2\pi {\rm Im} \tau \Delta_0)}
{{\rm det}( 2\pi {\rm Im} \tau \Delta_1)} \Big]^{\frac{1}{2}}
=(\frac{g}{(2\pi)^4\sqrt{g}})^{\frac{1}{2}} 
\left (2\pi{\rm Im} \tau\right)^{\frac{b_1 - 1}{2}} \hskip5pt 
\frac{{\rm det} \Delta_0}{{\rm det} \Delta_1^\frac{1}{2}}, 
\end{align}
where $b_1 =4$ is the dimension of the group $H^1(T^4)$. $\Delta_p 
= (d^\dagger d + d d^\dagger)_p$ is the Laplacian operator acting on 
the $p$-form. 
$g = {\rm det} G_{ij}$. So 
$\Delta_0= -G^{ij}\partial_i\partial_j$,
and ${\rm det} (\Delta_1) = {\rm det}(\Delta_0)^4$.
Thus 
{\begin{align}
Z^{PI}_{{\rm non-zero\, modes}}=\frac{g^{\frac{1}{4}}}{\sqrt{2\pi}} 
({\rm Im} \tau)^{\frac{3}{2}} \hskip5pt {\rm det} \Delta_0^{-1} .     
\label{sim}\end{align} 
The determinant can be computed
\begin{align}
{\rm det} \Delta_0^{-\frac{1}{2}} 
=\hskip1pt {\rm exp} \{-\frac{1}{2} 
{\rm tr}{\rm ln} A\},
\label{important-id}\end{align} 
\begin{align}
{\rm exp}\{ -\frac{1}{2} {\rm  tr {\rm ln} \Delta_0}\}  &= {\rm exp} \Big( -\frac{1}{2} {\rm  tr ln \big(  -G^{66} \partial_6^2 - 2G_{6\alpha} \partial_6 \partial_\alpha -G^{\alpha\beta} \partial_\alpha \partial_\beta   \big)} \Big) \cr
&=  {\rm exp} \Big( -\frac{1}{2} {\rm  \sum_{n_\alpha\ne \vec 0} 
\sum_{n_6} ln
\big(  \frac{1}{R_6^2} n_6^2 + 2 \frac{\gamma^\alpha}{R_6^2} n_\alpha n_6 
+ G^{\alpha\beta} n_\alpha n_\beta\big)} \Big) \cr
&=  {\rm exp} \Big( -\frac{1}{2} {\rm  \sum_{n_\alpha\ne\vec 0}
\sum_{n_6} 
ln \big(  \frac{1}{R_6^2} (n_6+ \gamma^\alpha n_\alpha)^2 
+ g^{\alpha\beta} n_\alpha n_\beta\big)} \Big). \cr
\end{align}
Let $\mu(E) \equiv \sum_{n_6} {\rm ln} \big( \frac{1}{R_6^2} 
(n_6 + \gamma^\alpha n_\alpha)^2 + E^2 \big)$, where $E^2 \equiv 
g^{\alpha\beta} n_\alpha n_\beta$, \;$\rho=2\pi R_6$,
\begin{align}
\frac{\partial \mu(E)}{\partial E} &= \sum_{n_6} \frac{2E}{ 
\frac{1}{R_6^2} (n_6 + \gamma^\alpha n_\alpha)^2 + E^2} = 
\frac{\rho \,{\rm sinh} (\rho E)}{{\rm cosh} (\rho E) - {\rm cos}
(2\pi \gamma^\alpha n_\alpha)}\cr
&= \partial_E {\rm ln} \big[ {\rm cosh} (\rho E) - {\rm cos}
(2\pi \gamma^\alpha n_\alpha)  \big].
\end{align}
After integration, we have
\begin{align}
\mu(E) = {\rm ln} \Big[  {\rm cosh} (\rho E) - {\rm cos} 
(2\pi \gamma^{\alpha} n_\alpha) \Big] + {\rm ln} \left( 
R_6^2\sqrt{2\over\pi}\right).
\end{align}
where the constant 
${\rm ln} \left( R_6^2\sqrt{2\over\pi}\right) $ 
maintains $SL(4, {\cal Z})$ invariance of the partition function. So,
\begin{align}
{{\rm det} \Delta_0}^{-\frac{1}{2}} = {\rm exp} \Big( -\frac{1}{2} 
{\rm tr {\rm ln} \Delta_0} \Big) 
&= e^{-\frac{1}{2} \sum_{n_\alpha\ne\vec 0} \mu(E)}\cr
&= \frac{(2\pi)^{\frac{1}{4}}}{R_6} \prod_{n_\alpha\in {\cal Z}^3\ne\vec 0} 
\frac{1}{\sqrt{2}\sqrt{{\rm cosh} (\rho E) - {\rm cos} (2\pi \gamma^{\alpha} 
n_\alpha)  }} \cr
&=  \frac{(2\pi)^{\frac{1}{4}}}{R_6} \prod_{n_\alpha \in {\cal Z}^3\ne \vec 0} 
\frac{e^{-\frac{\rho E}{2}}}{ 1-e^{-\rho E + 2\pi i \gamma^\alpha n_{\alpha}}}.
\label{last}\end{align}
Therefore, using (\ref{sim}), we have
\begin{align}
Z^{PI}_{\rm non-zero\, modes} 
=({\rm Im}\,\tau)^{\frac{3}{2}} 
\frac{ g^{\frac{1}{4}}}{R_6^2} \hskip2pt Z^{4d}_{osc},
\label{agrelationosc}\end{align}
which is (\ref{relationosc}).

Together with (\ref{relationzero}), the partition functions 
from the two quantizations agree but they factor differently into zero and 
oscillator modes,
\begin{align}
Z^{4d,Maxwell}=
Z^{4d}_{\rm zero\,modes} Z^{4d}_{\rm osc} = 
Z^{PI}_{\rm zero\,modes} Z^{PI}_{\rm non-zero\, modes}.
\label{allfour}\end{align}

\section{\bf $SL(2,{\cal Z})$ invariance of the 
$Z^{6d, chiral}$ and $Z^{4d, Maxwell}$ partition functions}

The S-duality group
$SL(2,{\cal Z})$ group has two generators $S$ and $T$ which act on the 
parameter $\tau$ to give
\begin{align}
S: \tau \rightarrow -\frac{1}{\tau}, \qquad  T: \tau \rightarrow \tau - 1.
\label{SL2Z}\end{align}
Since $\tau = \beta^2 + i {R_1\over R_2} = 
{\theta\over 2\pi} + i{4\pi\over e^2},$
the transformation $S$ corresponds to
\begin{align}
R_1\rightarrow R_1|\tau|^{-1}, \qquad R_2 \rightarrow R_2|\tau|, \qquad 
\beta^2 \rightarrow -|\tau|^{-2} \beta^2,
\label{Strans}\end{align}
and $T$ corresponds to 
\begin{align}
\beta^2 \rightarrow \beta^2 -1.
\label{Ttrans}\end{align}
Or equivalently 
\begin{align}
S:&\qquad {4\pi\over e^2}\rightarrow {4\pi\over e^2} |\tau|^{-2},\quad
\theta\rightarrow -\theta |\tau|^{-2}\cr
T:&\qquad \theta\rightarrow \theta - 2\pi,
\end{align}
which for $\theta = 0$ is the familiar electromagnetic duality transformation
${e^2\over 4\pi}\rightarrow  {4\pi\over e^2}$.
\vskip20pt

{\it $6d$ partition function}

The $6d$ chiral boson zero mode partition function (\ref{zmpf}),
\begin{align}
Z^{6d}_{\rm zero\, modes} =& \sum_{n_8,n_9,n_{10}} {\rm exp}
\{-
\frac{\pi R_6}{R_1 R_2}\sqrt{\tilde g}
g^{\alpha\alpha'}H_{12\alpha}H_{12\alpha'} \}\cr
&\hskip10pt
\cdot \sum_{n_7}{\rm exp} \{-\frac{\pi}{6}R_6R_1 R_2 \sqrt{\tilde g}
g^{\alpha\alpha'}g^{\beta\beta'}g^{\delta\delta'}H_{\alpha\beta\delta}H_{\alpha'\beta'\delta'}  -i\pi\gamma^{\alpha}\epsilon^{\gamma\beta\delta}H_{12\gamma}H_{\alpha\beta\delta}  \}\cr
&\cdot \hskip-6pt
\sum_{n_4,n_5,n_6} {\rm exp} \{-\frac{\pi}{2}R_6R_1 R_2\sqrt{\tilde g}(\frac{1}{R_2^2}+\frac{{\beta^2}^2}{{R_1}^2})g^{\alpha\alpha'}g^{\beta\beta'}H_{2\alpha\beta}H_{2\alpha'\beta'} \}\cr
&\hskip10pt\cdot\hskip-6pt
\sum_{n_1,n_2,n_3} {\rm exp} \{-\pi\frac{R_6 R_2}{R_1} \sqrt{\tilde g}
\beta^2 g^{\alpha\alpha'}g^{\beta\beta'}H_{1\alpha\beta}H_{2\alpha'\beta'}
+ i\pi\gamma^{\alpha}\epsilon^{\gamma\beta\delta} H_{1\gamma\beta}
H_{2\alpha\delta}\cr
&\hskip70pt
-\frac{\pi}{4}\frac{R_6 R_2}{R_1} \sqrt{\tilde g}(g^{\alpha\alpha'}
g^{\beta\beta'}-g^{\alpha\beta'}g^{\beta\alpha'})H_{1\alpha\beta}
H_{1\alpha'\beta'} \}\label{zmpfappb}\end{align}
where $H_{134} = n_1$,$H_{145} = n_2$, $H_{135} = n_3$, $H_{234} = n_4$,
$H_{245} = n_5$, $H_{235} = n_6$, $H_{345} = n_7$, $H_{123} = n_8$,
$H_{124} = n_9$, $H_{125} = n_{10}$,
is invariant under both $S$ and $T$.
To show the invariance using (\ref{Strans},\ref{Ttrans}) we group the exponents in (\ref{zmpfappb})
into two sets,
\begin{align}&
-\frac{\pi R_6}{R_1 R_2}\sqrt{\tilde g}
g^{\alpha\alpha'}H_{12\alpha}H_{12\alpha'}
-\frac{\pi}{6}R_6R_1 R_2\sqrt{\tilde g}g^{\alpha\alpha'}
g^{\beta\beta'}g^{\delta\delta'}H_{\alpha\beta\delta}H_{\alpha'\beta'\delta'}
-i\pi\gamma^{\alpha}\epsilon^{\gamma\beta\delta}H_{12\gamma}
H_{\alpha\beta\delta},
\label{irrelevantt}\end{align}
and
\begin{align}
&-\frac{\pi}{2}R_6R_1 R_2 \sqrt{\tilde g}( \frac{1}{R_2^2}+
\frac{{\beta^2}^2}{{R_1}^2})g^{\alpha\alpha'}g^{\beta\beta'}H_{2\alpha\beta}
H_{2\alpha'\beta'}
-\pi\frac{R_6}{R_1}\ R_2\sqrt{\tilde g} \beta^2 g^{\alpha\alpha'}g^{\beta\beta'}H_{1\alpha\beta}H_{2\alpha'\beta'} \cr
& + i\pi\gamma^{\alpha}\epsilon^{\gamma\beta\delta}
H_{1\gamma\beta} H_{2\alpha\delta}
-\frac{\pi}{2}\frac{R_6R_2}{R_1} \sqrt{\tilde g}\,g^{\alpha\alpha'}
g^{\beta\beta'}\,H_{1\alpha\beta}
H_{1\alpha'\beta'}.
\label{relevantt}\end{align}
(\ref{irrelevantt}) has no $\beta^2$ dependence and therefore is invariant
under $T$. (\ref{relevantt}) transforms
under $T$ to become
\begin{align}
&-\frac{\pi}{2}R_6R_1 R_2 \sqrt{\tilde g}( \frac{1}{R_2^2}+\frac{{\beta^2}^2}{{R_1}^2})g^{\alpha\alpha'}g^{\beta\beta'}H_{2\alpha\beta}H_{2\alpha'\beta'}
-\pi\frac{R_6}{R_1}\ R_2\sqrt{\tilde g} \beta^2 g^{\alpha\alpha'}g^{\beta\beta'}H_{1\alpha\beta}H_{2\alpha'\beta'} \cr & + i\pi\gamma^{\alpha}\epsilon^{\gamma\beta\delta}
H_{1\gamma\beta} H_{2\alpha\delta}
-\frac{\pi}{2}\frac{R_6}{R_1 R_2} \sqrt{\tilde g}\,g^{\alpha\alpha'}
g^{\beta\beta'}\,H_{1\alpha\beta}
H_{1\alpha'\beta'}\cr
&+ \pi\frac{R_6}{R_1} \sqrt{\tilde g} R_2
\beta^2 g^{\alpha\alpha'}g^{\beta\beta'} H_{2\alpha\beta} H_{2\alpha'\beta'}
- \frac{\pi}{2} \frac{R_6}{R_1}\sqrt{\tilde g} R_2 g^{\alpha\alpha'}
g^{\beta\beta'} H_{2\alpha\beta} H_{2\alpha'\beta'}
+ \pi\frac{R_6}{R_1} R_2\sqrt{\tilde g} g^{\alpha\alpha'} g^{\beta\beta'}
H_{1\alpha\beta} H_{2\alpha'\beta'},\cr
\end{align}
which is equivalent to (\ref{relevantt}) in the sum where we shift the integer
zero mode
field strength $H_{1\alpha\beta}$ to $H_{1\alpha\beta} - H_{2\alpha\beta}$.

Under $S$, 
we see (\ref{irrelevantt}) as a function of $R_1R_2$ is invariant,
and find (\ref{relevantt}) transforms to
\begin{align}
&-{\pi\over 2}
\frac{R_6R_2}{R_1}\sqrt{\tilde g} g^{\alpha\alpha'} g^{\beta\beta'}
H_{2\alpha\beta} H_{2\alpha'\beta'}
+ \pi\frac{R_6}{R_1} R_2 \sqrt{\tilde g}\beta^2g^{\alpha\alpha'}g^{\beta\beta'} H_{1\alpha\beta}H_{2\alpha'\beta'}\cr
& + i\pi\gamma^{\alpha}\epsilon^{\gamma\beta\delta}
H_{1\gamma\beta} H_{2\alpha\delta}
-\frac{\pi}{2} R_1 R_6 R_2 \sqrt{\tilde g} (g^{22} +
\frac{{\beta^2}^2}{R_1^2})g^{\alpha\alpha'} g^{\beta\beta'}
H_{1\alpha\beta}H_{1\alpha'\beta'}.
\end{align}
So by shifting the integer field strength tensors
$H_{1\alpha\beta}\rightarrow H_{2\alpha\beta}$ and
$H_{2\alpha\beta}\rightarrow -H_{1\alpha\beta}$,
the sum on (\ref{relevantt}) is left invariant by $S$.
Thus we have proved $SL(2,{\cal Z})$ invariance of the $6d$ zero mode partition
function (\ref{zmpf}), and that its factors $\epsilon$ and $Z^{4d}_{\rm
zero modes}$ in (\ref{zmpfsep}) are separately $SL(2,{\cal Z})$ invariant.

For the oscillator modes (\ref{ZHplusP6d}),
the only term that transforms in the sum and product is
\begin{align}
\tilde p^2 \equiv \frac{{p_1}^2}{{R_1}^2} + (g^{22}+\frac{{\beta^2}^2}{{R_1}^2})
p_2^2 + \frac{2\beta^2}{{R_1}^2}p_1p_2,
\end{align}
which is invariant under $T$ by shifting the momentum
$p_1\rightarrow p_1 + p_2$.
With the $S$ transformation, $\tilde p^2$ becomes
\begin{align}
{p_1}^2(g^{22} + \frac{{\beta^2}^2}{{R_1}^2})
+ \frac{1}{{R_1}^2}p_2^2 - \frac{2\beta^2}{{R_1}^2}p_1p_2,
\end{align}
and by also exchanging the momentum $p_1\rightarrow p_2$ and
$p_2\rightarrow -p_1$, the term remains the same.
So the 6d oscillator partition function (\ref{ZHplusP6d})
is $SL(2{\cal Z})$ invariant, which holds also for regularized vacuum
energy as given in (\ref{f6doscpf}).

\vskip15pt
{ \it $4d$ $U(1)$ partition function}

In the Hamiltonian formulation, 
$SL(2,{\cal Z})$ leaves invariant the $U(1)$ oscillator partition function 
(\ref{opf}), since it is independent of $e^2$ and $\theta$. 
We have also checked above, starting from $6d$, that
the zero mode $4d$ partition  function (\ref{zmpf4d}) is invariant.
Thus the $U(1)$ partition function from the Hamiltonian formalism is 
S-duality invariant.

The S-duality transformations on the corresponding quantities
in the path integral quantization can be derived from (\ref{relationzero}) and
(\ref{agrelationosc}). 
Since ${\rm Im}\,\tau\rightarrow {1\over |\tau|^2}{\rm Im}\tau$
under S, and is invariant under T, we have
\begin{align}
&Z^{4d}_{\rm zero\; modes}\longrightarrow Z^{4d}_{\rm zero \; modes},\qquad
Z^{PI}_{\rm zero\; modes} \longrightarrow |\tau|^{3}
Z^{PI}_{\rm zero \; modes}
\hskip35pt \hbox{under $S$}\cr
&Z^{4d}_{\rm zero\; modes} \longrightarrow Z^{4d}_{\rm zero \; modes},
\qquad
Z^{PI}_{\rm zero\; modes} \longrightarrow Z^{PI}_{\rm zero \; modes}
\qquad\hskip23pt
\hskip8pt\hbox{under $T$}
\end{align}
and
\begin{align}
& Z^{4d}_{\rm osc} \longrightarrow Z^{4d}_{\rm osc},\qquad
Z^{PI}_{\rm osc} \longrightarrow |\tau|^{-3} Z^{PI}_{\rm osc}
\hskip35pt \hbox{under $S$}\cr
& Z^{4d}_{\rm osc}\longrightarrow Z^{4d}_{\rm osc}, \hskip23pt
Z^{PI}_{\rm osc} \longrightarrow Z^{PI}_{\rm osc}
\hskip58pt \hbox{under $T$},
\end{align}
which is (\ref{agzmtransf}) and (\ref{agosctransf}).

\vskip15pt
{ \it A $2d$-$4d$ correspondence for $Z^{PI}_{\rm zero\, modes}$}}

We remark here that the zero mode contribution to the partition function for the 4d Maxwell field
is equivalent to the zero mode contribution to the partition function
for a 2d worldsheet action \cite{Verlinde},
\begin{align}
S= \int d^2\sigma [\sqrt{h} h^{\mu\nu} G_{\alpha\beta}\partial_\mu X^\alpha
\partial_\nu X^\beta + \epsilon^{\mu\nu} B_{\alpha\beta}\partial_\mu X^\alpha
\partial_\nu X^\beta ],\label{2daction}
\end{align}
where $1\le \mu,\nu\le 2$ and $3\le \alpha,\beta\le 5$.
\begin{align}
Z^{2d}_{\rm zero\, modes} &=
\sum_{(p_l,p_r)\in \Gamma_{3,3}} e^{i\pi\tau (p_L)^2
-i\pi\bar\tau (p_R)^2}
= \sum_{(p_l,p_r)\in \Gamma_{3,3}} e^{i{\theta\over 2}((p_L)^2 -
(p_r)^2) -{4\pi^2\over e^2} ((p_L)^2 + (p_r)^2)}\cr
&= \sum_{n_\alpha\in {\cal Z}^3, m^\beta\in  {\cal Z}^3}
e^{i\theta n_\alpha m^\alpha
-{4\pi^2\over e^2}(m^\alpha m^\beta G_{\alpha\beta}
+(n_\alpha - B_{\alpha\rho}m^\rho) G^{\alpha\beta} (n_\beta-B_{\beta\sigma}
m^\sigma))} \cr
&=Z^{PI}_{\rm zero\, modes}
\nonumber\end{align}
where $Z^{PI}_{\rm zero\, modes}$ is given in (\ref{f2}). 
The 2d metric is
$h_{11}= R_1^2 + R_2^2\beta^2\beta^2, \,h_{12}=-R_2^2\beta^2,$\hfill\break
$h_{22}=R_2^2,$
and $\tau = \beta^2 + i{R_1\over R_2} = {\theta\over 2\pi} + i{4\pi\over e^2}.$
The nine parameters of the moduli space ${O(3,3)\over O(3)\times O(3)}$
of the Lorentzian lattice $\Gamma_{3,3}$ are given by the
4d gauge theory metric as
\begin{align}
G_{\alpha\beta}&= {R_6\over \sqrt{\tilde g}} g_{\alpha\beta},\qquad
B_{\alpha\beta}= {\epsilon_{\alpha\beta\lambda} \gamma^\lambda
\over {\tilde g}}.
\nonumber\end{align}
The integers are identified with the Maxwell field components as
\begin{align}
\widetilde F_{6\alpha} = n_\alpha,\qquad
\widetilde F_{\alpha\beta} = {\epsilon_{\alpha\beta\rho} m^\rho
\over {\tilde g}}.
\nonumber\end{align}
The points $(p_{L\gamma},
p_{R\delta})$ on the
Lorentzian lattice $\Gamma_{d,d}$ are \cite{GPR}
\begin{align}
p_{L\gamma} &= n_\alpha e^{\ast\alpha}_\gamma + m^\alpha (B_{\alpha\beta}
+ G_{\alpha\beta} ) e^{\ast\beta}_\gamma,\qquad
p_{R\gamma} = n_\alpha e^{\ast\alpha}_\gamma + m^\alpha (B_{\alpha\beta}
- G_{\alpha\beta} ) e^{\ast\beta}_\gamma,\cr
p_L^2 &= \sum_{\gamma =1}^d p_{L\gamma} p_{L\gamma},\quad
p_R^2 = \sum_{\gamma =1}^d p_{R\gamma} p_{R\gamma},\quad
\sum_{\gamma =1}^d e^{\ast\alpha}_\gamma e^{\ast\beta}_\gamma =
{1\over 2} G^{\alpha\beta},\quad p_L^2 - p_R^2 = 2 n_\alpha m^\alpha,\cr
p_L^2 &+ p_R^2 = m^\alpha m^\beta G_{\alpha\beta}
+(n_\alpha - B_{\alpha\rho}m^\rho) G^{\alpha\beta} (n_\beta-B_{\beta\sigma}
m^\sigma),
\nonumber\end{align}
for $1\le \alpha,\beta,\gamma,\delta\le d$.

However, the non-zero mode partition function of the 2d theory (\ref{2daction})
\begin{align}
Z^{2d}_{\rm non-zero\,modes}&= \left(\eta(\tau)\bar\eta(\bar\tau)\right)^{-3}
\label{2dnonzeromodes}\end{align}
is not the $4d$ non-zero mode partition function (\ref{agrelationosc}),
although they both transform in the same way under the $SL(2,{\cal Z})$
duality transformation. Indeed (\ref{agrelationosc}) is more naturally
described by a 2d scalar theory with massless and massive modes on 
a two-torus in the directions 3 and 6, as we show in Appendix D.

\section{Regularization of the vacuum energy for $4d$ Maxwell theory}
The sum in (\ref{opf}) is divergent. 
We regularize the vacuum energy following \cite{DN},\cite{DS}.
For $<H> = {1\over 2}\sum_{p_\alpha\in {\cal Z}^3}
\sqrt{g^{\alpha\beta}p_\alpha p_\beta}$, 
the $SL(3,{\cal Z})$ invariant regularized vacuum energy becomes 
\begin{align}
<H> = -{1\over 4\pi^3}\sqrt{\tilde g} \sum_{n^\alpha \in {\cal Z}^3\ne 0}
{1\over (g_{\alpha\beta}n^\alpha n^\beta)^{2}} = -4\pi \sqrt{\tilde g}
\sum_{{\vec n}\in{\cal Z}^3\ne 0} 
{1\over |2\pi {\vec n}|^4}.\label{still}
\end{align}

For the proof of $SL(4, {\cal Z})$ invariance in Appendix D,
it is also useful to write the regularized sum (\ref{still}), 
as 
\begin{align}
<H> &= \sum_{p_\perp\in {\cal Z}^2} <H>_{p_\perp} 
= \; <H>_{p_\perp=0} \; + \sum_{p_\perp\in {\cal Z}^2\ne 0}<H>_{p_\perp},
\label{hsplit}\end{align}
where $p_\perp =p_a \in {\cal Z}^2,\; a=4,5,$ and 
\begin{align}
<H>_{p_\perp=0} &= {1\over 2} \sum_{p_3\in {\cal Z}} \sqrt{g^{3 3}p_3p_3}
= {1\over R_3} \sum_{n=1}^\infty n =  {1\over R_3}\zeta(-1)
= -{1\over 12 R_3};\cr
<H>_{p_\perp\ne 0}&= |p_\perp|^2
R_3 \sum_{n=1}^\infty\cos(p_a \kappa^a 2\pi n)
\big [ K_2(2\pi n R_3 |p_\perp|) - K_0(2\pi n R_3 |p_\perp|)\big].
\label{C3}\end{align}
$ |p_\perp| = \sqrt{p_ap_b{\tilde g}^{ab}}$, 
using the 2d inverse metric as defined in Appendix D.

\section{$SL(4,\cal Z)$ invariance of 
$Z^{4d, {\rm Maxwell}}$ and $Z^{6d,{\rm chiral}}$}

{\it Rewriting the $4d$ metric (3,4,5,6)}

From (\ref{metric1}) the metric on the four-torus, 
for $\alpha, \beta = 3,4,5$, is
\begin{align}
G_{\alpha\beta}&= g_{\alpha\beta},\qquad G_{\alpha 6}
= - g_{\alpha\beta}\gamma^\beta,\qquad
G_{66}= R_6^2 + g_{\alpha\beta}\gamma^\alpha\gamma^\beta.
\end{align}
We can rewrite this metric using $a, b = 4,5$,
\begin{align}
g_{33}& \equiv R_3^2 + g_{ab}\kappa^a\kappa^b,
\qquad g_{a 3} \equiv -g_{ab} \kappa^b,\qquad
g_{ab}\equiv g_{ab},\qquad
(\gamma^3)\kappa^a - \gamma^a \equiv -\widetilde\gamma^a,\\
\cr
G_{33}&=  R_3^2 + g_{ab}\kappa^a \kappa^b,
\qquad G_{36}= - (\gamma^3) R_3^2 + g_{ab}\kappa^b
\widetilde\gamma^a,\qquad G_{3a} = -g_{ab} \kappa^b,\cr G_{ab} &= g_{ab},
\hskip70pt
G_{a 6} = - g_{ab} \widetilde\gamma^b,
\hskip40pt G_{66} = R_6^2 + (\gamma^3)^2 R_3^2 + g_{ab}
\,\widetilde\gamma^a \widetilde\gamma^b.
\label{our5dmetric}\end{align}

The $3d$ inverse of $g_{\alpha \beta}$ is
\begin{align}
g^{ab} = \widetilde g^{a b} + {\kappa^a \kappa^b \over
R_3^2},\qquad g^{a 3} = {\kappa^a\over R_3^2},\qquad
g^{33}={1\over R_3^2},
\end{align} where $\widetilde g^{a b}$ is the $2d$
inverse of $g_{a b}$.
$$ g \equiv \det G_{ij} = R_6^2 \; \det g_{\alpha\beta} 
\equiv  R_6^2 \; \widetilde g  = R_6^2 R_3^2 \; \det g_{ab}  
\equiv  R_6^2 R_3^2\; \bar g .$$
The line element can be written as
\begin{align}
ds^2 &= R_6^2 (d\theta^6)^2 + \sum_{\alpha, \beta=3,4,5} g_{\alpha\beta}
(d\theta^\alpha -\gamma^\alpha d\theta^6)(d\theta^\beta - \gamma^\beta d\theta^6)\cr
&= R_3^2 (d\theta^3 - (\gamma^3) d\theta^6)^2 + R_6^2 (d\theta^6)^2 \cr
&\hskip15pt
+ \sum_{a,b = 4,5} g_{ab}(d\theta^a
-\widetilde\gamma^a d\theta^6 - \kappa^a d\theta^3)\,
(d\theta^b
-\widetilde\gamma^b d\theta^6 - \kappa^b d\theta^3).
\label{5dlineelement}\end{align}
We define \begin{align}\widetilde \tau \equiv \gamma^3 + i {R_6\over R_3}.
\label{5dtau}\end{align}
The 4d inverse is
\begin{align}
&\widetilde G_4^{33} = {|\widetilde\tau|^2\over R_6^2} = \widetilde G_4^{66}
|\widetilde\tau|^2,\qquad \widetilde G_4^{66} = {1\over R^2_6},\qquad
\widetilde G_4^{36} = {\gamma^3\over R_6^2}, \qquad
\widetilde G_4^{3a} = {\kappa^a |\widetilde\tau|^2\over  R_6^2}
+ {\gamma^3\widetilde\gamma^a\over R_6^2},\cr
&\widetilde G_4^{ab} = \widetilde g^{ab} +
{\kappa^a\kappa^b\over R_6^2} |\widetilde\tau|^2
+ {\widetilde\gamma^a \widetilde\gamma^b\over R_6^2}
+ {\gamma^3(\widetilde\gamma^a\kappa^b +
\kappa^a\widetilde\gamma^b )\over R_6^2},\qquad
\widetilde G_4^{6a} ={\gamma^a\over R_6^2}
= {\gamma^3\kappa^a + \widetilde \gamma^a
\over R_6^2}.
\label{invalpha}\end{align}

{\it Generators of $GL(n,{\cal Z})$}

The $GL(n,{\cal Z})$ unimodular group can be generated by three
matrices \cite{Coxeter}. For $GL(4,{\cal Z})$ these can be taken to be
$U_1, U_2$ and $U_3$,
\begin{align}
U_1 &= \left (\begin{matrix}
0&1&0&0\cr
0&0&1&0\cr
0&0&0&1\cr
1&0&0&0\end{matrix}\right)\,;\qquad
U_2 = \left (\begin{matrix}
1&0&0&0\cr
1&1&0&0\cr
0&0&1&0\cr
0&0&0&1\cr
\end{matrix}\right)\,; \qquad
U_3 = \left (\begin{matrix}
-1&0&0&0\cr
0&1&0&0\cr
0&0&1&0\cr
0&0&0&1\cr
\end{matrix}\right),
\end{align}

so that every matrix $M$ in
$GL(4,{\cal Z})$ can be written
as a product $U_1^{n_1} U_2^{n_2} U_3^{n_3}U_1^{n_4} U_2^{n_5} U_3^{n_6}\dots$,
for integers $n_i$.
Matrices $U_1$, $U_2$ and $U_3$ act on the basis vectors of the
four-torus $\vec\alpha_{i}$ where $\vec\alpha_{i}
\cdot\vec\alpha_{j}\equiv
\alpha_{i}^{k} \alpha_{j}^{l} G_{kl}
= G_{ij}$,
\begin{align}
\vec\alpha_{3}&= (1,0,0,0)\cr
\vec\alpha_{6}&= (0,1,0,0)\cr
\vec\alpha_{4}&= (0,0,1,0)\cr
\vec\alpha_{5}&= (0,0,0,1).
\end{align}
For our metric (\ref{our5dmetric}), the $U_2$ transformation
\begin{align}\left(\begin{matrix}\vec\alpha'_3\cr
\vec\alpha'_6\cr
\vec\alpha'_4\cr
\vec\alpha'_5
\end{matrix}\right)
=U_2\left(\begin{matrix}\vec\alpha_3\cr
\vec\alpha_6\cr
\vec\alpha_4\cr
\vec\alpha_5\end{matrix}\right) =
\left (\begin{matrix}
1&0&0&0\cr
1&1&0&0\cr
0&0&1&0\cr
0&0&0&1\cr
\end{matrix}\right)
\end{align} results in
$\vec\alpha'_{3}
\cdot\vec\alpha'_{3}\equiv
{\alpha'}^{i}_3 {\alpha'}^{j}_3 G_{ij}
= G_{33} = G'_{33}, \;\;$$\vec\alpha'_{3}
\cdot\vec\alpha'_{6}\equiv
{\alpha'}^{i}_{3}{\alpha'}^{j}_{6} G_{ij}
= G_{33} + G_{36} = G'_{36}$, etc.
So $U_2$ corresponds to
\begin{align} R_3\rightarrow R_3 , \; R_6  \rightarrow R_6, \;\gamma^3
\rightarrow \gamma^3 - 1, \;\kappa^a\rightarrow \kappa^a,\;
\widetilde\gamma^a \rightarrow \widetilde\gamma^a
+ \kappa^a, \;g_{ab}\rightarrow
g_{ab},\label{U2transf}\end{align}
or equivalently
\begin{align}
R_6\rightarrow R_6, \; \gamma^3\rightarrow \gamma^3 -1,\;
g_{\alpha\beta}\rightarrow g_{\alpha\beta},\; \gamma^a \rightarrow\gamma^a,
\label{transfu2}\end{align}
which leaves invariant the line element (\ref{5dlineelement})
if $d\theta^3\nobreak\rightarrow d\theta^3 - d\theta^6, \,
d\theta^6 \rightarrow d\theta^6,\,
d\theta^a \rightarrow d\theta^a.$
$U_2$ is the generalization of the usual $\widetilde\tau\rightarrow
\widetilde\tau - 1$
modular transformation.
The 3d inverse metric $g^{\alpha\beta} \equiv \{g^{ab}, g^{a 3}, g^{33}\}$
does not change under $U_2$.
It is easily checked that $U_2$ is an invariance of the 4d Maxwell
partition function (\ref{wpf4}) as well as the
$6d$ chiral boson partition function (\ref{f6doscpf}). It leaves the zero mode
and oscillator contributions invariant separately. 

The other generator, $U_1$ is related to the 
$SL(2,{\cal Z})$ transformation $\widetilde\tau\rightarrow 
-(\widetilde\tau)^{-1}$ that we discuss as follows:
\begin{align}
U_1 = U' M_3
\label{U1}\end{align}
where $M_3$ is a $GL(3,{\cal Z})$ transformation given by
\begin {align}
M_3 &= \left (\begin{matrix}
0&0&-1&0\cr
0&1&0&0\cr
0&0&0&1\cr
1&0&0&0\cr
\end{matrix}\right)\end{align}

and $U'$ is the matrix corresponding to the transformation on the
metric parameters (\ref{5dmod}),
\begin {align}
U' &= \left (\begin{matrix}
0&1&0&0\cr
-1&0&0&0\cr
0&0&1&0\cr
0&0&0&1\cr
\end{matrix}\right).
\end{align}
Under $U'$, the metric parameters transform as
\begin{align}
R_3&\rightarrow R_3 |\widetilde\tau|,\quad
R_6\rightarrow R_6  |\widetilde\tau|^{-1},\quad
\gamma^3\rightarrow -\gamma^3 |\widetilde\tau|^{-2},\quad
\kappa^a\rightarrow\widetilde\gamma^a,\quad
\widetilde\gamma^a\rightarrow - \kappa^a,\quad
g_{ab}\rightarrow g_{ab}.\cr
\widetilde\tau&\rightarrow -{1\over \widetilde \tau}.
\qquad\qquad\hbox{Or equivalently,}\cr
G_{ab}&\rightarrow G_{ab},\quad
G_{a 3}\rightarrow G_{a 6},\quad
G_{a 6}\rightarrow -G_{a 3},\quad
G_{33}\rightarrow G_{66},\quad
G_{66}\rightarrow G_{33},\quad
G_{36}\rightarrow - G_{36},\cr
\widetilde G_4^{ab} &\rightarrow \widetilde G_4^{ab},
\quad \widetilde G_4^{a 3}\rightarrow \widetilde G_4^{a 6},
\quad \widetilde G_4^{a 6}\rightarrow -\widetilde G_4^{a 3},
\quad \widetilde G_4^{3 3}\rightarrow {\widetilde G_4^{3 3}
\over |\widetilde\tau|^2},
\quad \widetilde G_4^{66}\rightarrow |\widetilde\tau|^2 \widetilde G_4^{66},
\quad \widetilde G_4^{36}\rightarrow
-\widetilde G_4^{36},
\label{5dmod}\end{align}\vskip-16pt
where $4\le a,b \le 5,$ and
\begin{align}
\widetilde \tau&\equiv \gamma^3 + i{R_6\over R_3},\qquad |\widetilde\tau|^2 =
(\gamma^3)^2 + {R^2_6\over R_3^2}.
\label{newtau}\end{align}
The transformation (\ref{5dmod}) leaves invariant the line element
(\ref{5dlineelement}) when $d\theta^3\rightarrow d\theta^6,\hfill\break
d\theta^6\rightarrow -d\theta^3,\,$ $d\theta^a\rightarrow d\theta^a.$
The generators have the property
$\det U_1 =-1, \det U_2=1,\hfill\break\det U_3=-1,\,
\det U'=1,\,\det M_3 = -1.$

Under $M_3$, the metric parameters transform as
\begin{align}
&R_6\rightarrow R_6,\qquad \gamma^3\rightarrow -\gamma^4,\quad
\gamma^a\rightarrow \gamma^{a+1},\quad
g_{ab}\rightarrow g_{a + 1, b + 1},\quad
g_{a 3}\rightarrow - g_{a + 1, 4},\quad
g_{3 3}\rightarrow g_{44},\cr
&g^{ab} \rightarrow g^{ a + 1, b + 1},\quad
g^{a 3}\rightarrow - g^{a + 1,  4},\quad g^{3 3}\rightarrow g^{4 4},
\quad \det g_{\alpha\beta}= {\tilde g}, 
\quad {\tilde g} \rightarrow {\tilde g}.
\qquad \hbox{Or equivalently,}\cr\cr
&G_{a b}\rightarrow G_{a + 1, b + 1},\quad
G_{a 3}\rightarrow -G_{ a + 1 , 4},\quad
G_{a 6}\rightarrow G_{ a + 1 , 6},\quad
G_{ 3 3 }\rightarrow G_{4 4},\quad
G_{66}\rightarrow G_{66},\quad G_{36}\rightarrow -G_{46},\cr
&\widetilde G_4^{ab}\rightarrow \widetilde G_5^{a + 1, b +1},
\quad \widetilde G_4^{a 3}\rightarrow -\widetilde G_5^{ a + 1, 4},
\quad\widetilde G_4^{a 6}\rightarrow \widetilde G_4^{  a + 1, 6},\quad
\widetilde G_4^{33}\rightarrow \widetilde G_4^{4 4},\quad
\widetilde G_4^{36}\rightarrow -\widetilde G_4^{4 6},\quad
\widetilde G_4^{66}\rightarrow \widetilde G_4^{66},\cr
&\det \widetilde G_4 = R_6 \;{\tilde g}, \qquad \det \widetilde G_4
\rightarrow \det \widetilde G_4,\label{M3}\end{align}
where $4\le a, b\le 5$, and $a +1 \equiv 3$ for $a =5$.
We see that $M_3$ takes ${Z}^{4d}_{\rm zero\, 
modes}$ to its complex conjugate as follows.
Letting the $M_3$ transformation (\ref{M3}) act on
(\ref{zmpf4d}), we find that the three subterms in the exponent
\begin{align}
&- {e^2\over 8} R_6\,\sqrt{\tilde g} ({\theta^2\over 4\pi^2}
+ {16\pi^2\over e^4}) \Big(
g^{aa'} g^{bb'}
\tilde F_{ab} \tilde F_{a'b'}
+ 4 g^{aa'} g^{b3}
\tilde F_{ab} \tilde F_{a'3}
+  2 g^{aa'} g^{33}
\tilde F_{a 3} \tilde F_{a' 3}
- 2 g^{a 3} g^{a' 3}
\tilde F_{a3} \tilde F_{a'3}\Big),\cr
&- {e^2 R_6\over 4\sqrt{\tilde g}}\,
{\tilde\Pi}^\alpha g_{\alpha\beta} {\tilde \Pi}^\beta,\cr
& -{\theta e^2 R_6\over 8\pi^2 \sqrt{\tilde g}}
g_{\alpha\beta} \epsilon^{\alpha\gamma\delta} 
{\tilde F}_{\gamma\delta} {\tilde \Pi}^{\beta},\end{align}\vskip-20pt
are separately invariant under (\ref{M3}) if we replace the integers  ${\widetilde F}_{\alpha\beta}\in {\cal Z}^3, {\tilde \Pi^\alpha}\in {\cal Z}^3$ by
\begin{align}
&{\tilde F}_{ab}\rightarrow {\tilde F}_{a+1, b+1},
\quad {\tilde F}_{a 3}\rightarrow - {\tilde F}_{a+1, 4},
\quad {\tilde \Pi}^3\rightarrow  {\tilde\Pi}^4,
\qquad {\tilde \Pi}^a \rightarrow -{\tilde \Pi}^{a + 1}.
\label{abnormalfieldshift}\end{align}
However, acted on by $M_3$ with the field shift 
(\ref{abnormalfieldshift}), the term
\begin{align}
2\pi i \gamma^{\alpha} \tilde\Pi^\beta {\widetilde F}_{\alpha\beta} \rightarrow -2\pi i \gamma^{\alpha} \tilde\Pi^\beta {\widetilde F}_{\alpha\beta} 
\end{align}
changes sign. Thus we have
\begin{align}
M_3: \hskip40pt  Z^{4d}_{\rm {zero\; modes}} 
\rightarrow Z^{{4d}\hskip5pt *}_{\rm {zero\; modes}} 
\label{M3transzero}\end{align}

{\it The action of $U'$ on ${Z}^{4d}_{\rm zero\; modes}$}

Next we show that under $U'$, ${Z}^{4d}_{\rm zero\, modes}$ transforms to 
$|\widetilde\tau|^2 \,  {Z}^{4d}_{\rm zero\, modes}$.
From (\ref{f2}) and (\ref{relationzero}), we have
\begin{align}
Z_{\rm zero \; modes} ^{4d} = ({4\pi\over e^2})^{-\frac{3}{2}} 
\frac{\tilde g^{\frac{1}{4}}}{R_6^{\frac{3}{2}}} 
\sum_{\tilde F_{ij} \in {\cal Z}^6} 
{\rm exp} \{- {2\pi^2\over e^2} R_6 \sqrt{\tilde g} g^{ij} g^{i'j'} 
\tilde F_{ii'} {\tilde F}_{jj'}- {i \over 2}\theta\epsilon^{\alpha\beta\gamma} 
{\tilde F}_{6\alpha} {\tilde F}_{\beta\gamma} \}
\label{transition},\end{align}
from which it will be easy to see how it transforms under the 
$U'$ transformation. Under $U'$ from (\ref{5dmod}), 
the coefficient transforms as 
\begin{align}
U':\qquad({4\pi\over e^2})^{-\frac{3}{2}} \frac{\tilde g^{\frac{1}{4}}}
{R_6^{\frac{3}{2}}}\rightarrow({4\pi\over e^2})^{-\frac{3}{2}} 
\frac{\tilde g^{\frac{1}{4}}}{R_6^{\frac{3}{2}}}\;|\widetilde\tau|^2.
\end{align}
The Euclidean action for the zero mode computation is invariant under $U'$, 
as we show next by first summing $ i = \{3, a, 6\},$ with
$4\le a \le 5.$
\begin{align}
& - {2\pi^2 R_6\sqrt{\tilde g}\over e^2}
\frac{R_1}{R_2} g^{ij} g^{i'j'} \tilde F_{ii'} {\tilde F}_{jj'}\cr
& =- {2\pi^2 R_6\sqrt{\tilde g}\over e^2}\Big(
\widetilde G_4^{aa'} \widetilde G_4^{bb'}
{\widetilde F}_{ab} {\tilde F}_{a'b'}
+ 4 \widetilde G_4^{aa'} \widetilde G_4^{b 3}
{\tilde F}_{ab} {\tilde F}_{a' 3}
+ 4 \widetilde G_4^{aa'} \widetilde G_4^{b 6}
{\tilde F}_{ab} {\widetilde F}_{a'6}
+ 2 \widetilde G_4^{aa'} \widetilde G_4^{33}
{\tilde F}_{a 3}{\tilde F}_{a'3}
\cr 
&\hskip80pt - 2 {\widetilde G}_4^{a 3} {\widetilde G}_4^{a' 3}
{\widetilde F}_{a 3} {\widetilde F}_{a' 3}
+ 4  \widetilde G_4^{aa'} \widetilde G_4^{36}
{\tilde F}_{a 3} {\widetilde F}_{a' 6}
-  4 \widetilde G_4^{a 6} \widetilde G_4^{a' 3}
{\tilde F}_{a 3} {\tilde F}_{a' 6}
+ 4 \widetilde G_4^{a 3} \widetilde G_4^{b 6}
{\tilde F}_{a b} {\tilde F}_{36}
\cr
&\hskip80pt 
+ 2  \widetilde G_4^{aa'} \widetilde G_4^{66}
{\tilde F}_{a 6 } {\tilde F}_{a' 6}
-  2 \widetilde G_4^{a 6} \widetilde G_4^{a' 6}
{\tilde F}_{a 6} {\tilde F}_{a' 6}
+ 4 \widetilde G_4^{a 3} \widetilde G_4^{36}
{\tilde F}_{a 3} {\tilde F}_{36}
- 4 \widetilde G_4^{a 6} \widetilde G_4^{33}
{\tilde F}_{a 3} {\tilde F}_{36} \cr
&\hskip80pt + 4 \widetilde G_4^{a 3} \widetilde G_4^{66}
{\tilde F}_{a 6} {\tilde F}_{36}
- 4  \widetilde G_4^{a 6} \widetilde G_4^{36}
{\tilde F}_{a 6} {\tilde F}_{36}
- 2  \widetilde G_4^{36} \widetilde G_4^{36}
{\tilde F}_{36} {\tilde F}_{36}
+ 2  \widetilde G_4^{33} \widetilde G_4^{66}
{\tilde F}_{36} {\tilde F}_{36}\Big).
\cr\label{zmalpha}\end{align}
Letting the $U'$ transformation (\ref{5dmod}) act on (\ref{zmalpha}),
we see the first term in the exponent of (\ref{transition}) changes to  

\begin{align}
&-{2\pi^2 R_6\sqrt{\widetilde g}\over e^2}\Big(
\widetilde G_4^{a a'} \widetilde G_4^{b b'}
{\tilde F}_{a b} {\tilde F}_{a' b'}
+ 4 \widetilde G_4^{a a'} \widetilde G_4^{b 6}
{\tilde F}_{a b} {\tilde F}_{a 3}
- 4 \widetilde G_4^{a a'} \widetilde G_4^{b 3}
{\tilde F}_{a b} {\tilde F}_{a' 6}
+ {2\over |\widetilde\tau|^2} 
\widetilde G_4^{a a'} \widetilde G_4^{33}
{\tilde F}_{a 3} {\tilde F}_{a' 3} 
\cr
&\hskip80pt 
- 2 \widetilde G_4^{a 6} \widetilde G_4^{a' 6}
{\tilde F}_{a 3} {\tilde F}_{a' 3}
- 4  \widetilde G_4^{a a'} \widetilde G_4^{3 6}
{\tilde F}_{a 3} {\tilde F}_{a' 6}
+  4 \widetilde G_4^{a 3} \widetilde G_4^{a' 6}
{\tilde F}_{a 3} {\tilde F}_{a' 6}
- 4 \widetilde G_5^{\alpha 6} \widetilde G_5^{\alpha' 3}
{\tilde F}_{a a'} {\tilde F}_{36}\cr
&\hskip80pt 
+ 2|\widetilde\tau|^2\widetilde G_4^{a a'} \widetilde G_4^{66}
F_{a 6 } F_{a' 6}
-  2 \widetilde G_4^{a 6} \widetilde G_4^{a' 3}
{\tilde F}_{a 6} {\tilde F}_{a' 6}
- 4 \widetilde G_4^{a 6} \widetilde G_4^{36}
{\tilde F}_{a 3} {\tilde F}_{36}
+ {4\over |\widetilde\tau|^2}
 \widetilde G_4^{a 3} \widetilde G_4^{33}
{\tilde F}_{a 3} {\tilde F}_{36}\cr
&\hskip80pt + 4 |\widetilde\tau|^2
\widetilde G_4^{a 6} \widetilde G_4^{66}
{\tilde F}_{a 6} {\tilde F}_{36}  
- 4  \widetilde G_4^{a 3} \widetilde G_4^{36}
{\tilde F}_{a 6} {\tilde F}_{36}
- 2  \widetilde G_4^{3 6} \widetilde G_4^{3 6} {\tilde F}_{36} {\tilde F}_{36}
+ 2  \widetilde G_4^{3 3} \widetilde G_4^{66} {\tilde F}_{36} {\tilde F}_{3 6}\Big).
\cr \label{zmtransf}\end{align}
The second term in the exponential of (\ref{transition}) is a topological 
term, and is left invariant under the action of $U'$ by inspection.
If we replace the integers ${\widetilde F}_{3a} \rightarrow 
{\widetilde F_{6a}}$ and ${\widetilde F}_{a6} \rightarrow 
- {\widetilde F}_{a3}$, the two terms are left invariant, so
the sum 
\begin{align}
\sum_{{\tilde F}_{ij}\in {\cal Z}^6 }\; 
e^{- {2\pi^2\sqrt{g}\over e^2} g^{ij} g^{i'j'} \tilde F_{ii'} 
{\tilde F}_{jj'} + i{\theta\over 2}\epsilon^{\alpha\beta\gamma} 
{\tilde F}_{6\alpha} {\tilde F}_{\beta\gamma} }
\end{align} is invariant. 
Thus we have shown that 
under the $U'$ transformation (\ref{5dmod}),
\begin{align}
{Z}^{4d}_{\rm zero\, modes} (R_3|\widetilde \tau|, R_6
|\widetilde \tau|^{-1}, g_{a b}, -\gamma^3
|\widetilde \tau|^{-2}, \widetilde\gamma^a, -\kappa^a)
= |\widetilde \tau|^2\; {Z}^{4d}_{\rm zero\, modes} (R_3, R_6,
g_{a b}, \gamma^3,
\kappa^a, \widetilde\gamma^a).
\label{noma}\end{align} 
Also from (\ref{transition}), we can write (\ref{M3transzero}) as
\begin{align}
M_3: \hskip40pt  Z^{4d}_{\rm zero\; modes} (e^2,\theta, G_{ij}) 
\rightarrow Z^{4d}_{\rm zero\; modes} (e^2, -\theta, G_{ij}) .
\label{M3another}\end{align}
and thus under the $GL(4,{\cal Z})$ generator $U_1$,
\begin{align}
{Z}^{4d}_{\rm zero\, modes} \rightarrow |\widetilde\tau|^2 \, \big( {Z}^{4d}_{\rm zero\, modes}\big)^{*}.
\end{align}
The residual factor $|{\widetilde \tau}|^2$ is sometimes referred to as an $SL(2, {\cal Z})$ anomaly of the 
zero mode partition function, because $U'$ includes the 
$\widetilde\tau\rightarrow -{1\over \widetilde \tau}$ transformation.
Finally we will show how this anomaly is canceled by the 
oscillator contribution.

Under $U_3$, the metric parameters transform as
\begin{align}
&R_6\rightarrow R_6,\qquad \gamma^3\rightarrow -\gamma^3,\quad
\gamma^a\rightarrow \gamma^{a},\quad
g_{ab}\rightarrow g_{a b},\quad
g_{a 3}\rightarrow - g_{a3},\quad
g_{3 3}\rightarrow g_{33},\cr
&g^{ab} \rightarrow g^{a b},\quad
g^{a 3}\rightarrow - g^{a3},\quad g^{3 3}\rightarrow g^{33},
\quad \det g_{\alpha\beta}= {\tilde g}, \quad {\tilde g} \rightarrow {\tilde g}.
\qquad \hbox{Or equivalently,}\cr\cr
&G_{a b}\rightarrow G_{a b},\quad
G_{a 3}\rightarrow -G_{ a 3},\quad
G_{a 6}\rightarrow G_{a 6},\quad
G_{33}\rightarrow G_{3 3},\quad
G_{66}\rightarrow G_{66},\quad G_{36}\rightarrow -G_{36},\cr
&\widetilde G_4^{ab}\rightarrow \widetilde G_4^{a b},
\quad \widetilde G_4^{a 3}\rightarrow -\widetilde G_4^{ a 3},
\quad\widetilde G_4^{a 6}\rightarrow \widetilde G_4^{  a 6},\quad
\widetilde G_4^{33}\rightarrow \widetilde G_4^{33},\quad
\widetilde G_4^{36}\rightarrow -\widetilde G_4^{36},\quad
\widetilde G_4^{66}\rightarrow \widetilde G_4^{66},\cr
&\det \widetilde G_4 = R_6 \;{\tilde g}, \qquad \det \widetilde G_4
\rightarrow \det \widetilde G_4,\label{U3}\end{align}
where $4\le a, b\le 5$ and $\widetilde G^{\alpha\beta}$ is the 3d inverse.
We can check that ${Z}^{4d}_{\rm zero\, modes}$ becomes its complex conjugate under $U_3$
given in (\ref{U3}) as follows.
Letting the $U_3$ transformation (\ref{U3}) act on
(\ref{zmpf4d}), we find that three of the terms in the exponent
\begin{align}
&-{e^2 R_6\sqrt{\tilde g}\over 8}
({\theta^2\over 4\pi^2} + {16\pi^2\over e^4})
\Big(
g^{aa'} g^{bb'}
F_{ab} F_{bb'}
+ 4 g^{aa'} g^{b3}
F_{ab} F_{a'3}
+  2 g^{aa'} g^{33}
F_{a 3} F_{a' 3}
- 2 g^{a 3} g^{a' 3}
F_{a3} F_{a'3}\Big),\cr
&-{e^2 R_6\over 4\sqrt{\tilde g}}
{\tilde\Pi}^\alpha g_{\alpha\beta} {\tilde \Pi}^\beta,\cr
& -{\theta e^2 R_6\over 8\pi\sqrt{\tilde g}}
g_{\alpha\beta} \epsilon^{\alpha\gamma\delta} {\tilde F}_{\gamma\delta} {\tilde \Pi}^{\beta},
\end{align}
are separately invariant under  (\ref{M3}), if we replace the
the integers ${\widetilde F}_{\alpha\beta}\in {\cal Z}^3, 
{\tilde \Pi^\alpha}\in {\cal Z}^3$ by
\begin{align}
&{\widetilde F}_{ab}\rightarrow {\widetilde F}_{a b},
\quad {\tilde F}_{a 3}\rightarrow - {\tilde F}_{a 3},
\quad {\tilde \Pi}^3\rightarrow {\tilde\Pi}^3,\qquad {\tilde \Pi}^a \rightarrow - {\tilde \Pi}^{a},
\label{U3fieldshift}\end{align}
However the subterm
\begin{align}
2 \pi i \gamma^\alpha {\tilde \Pi}^\beta \tilde F_{\alpha\beta} \rightarrow -2 \pi i \gamma^\alpha {\tilde \Pi}^\beta \tilde F_{\alpha\beta}
\end{align}
acted by $U_3$ with the field shift in (\ref{U3fieldshift}). Therefore 
the zero mode partition function goes to its complex conjugate
under $U_3$.
 
\bigskip
{\it Appropriate generators for $SL(4, {\cal Z})$}

We claim that $U_1^2$, $U_2$ and $U_1 U_3$ generate the group $SL(4, \cal Z)$ since $GL(n, {\cal Z})$ is generated by $U_1$, $U_2$ and $U_3$ or alternatively 
$R_1 = U_1$, $R_2 = U_3^{-1} U_2$ and $R_3 = U_3$, i.e., any element in $GL(n, {\cal Z})$ $U$ can be written as
\begin{align}
U = {R_1}^{n_1} {R_2}^{n_2} {R_3}^{n_3} {R_1}^{n_4} {R_2}^{n_5} 
{R_3}^{n_6} ... .
\end{align}
It is understood that $SL(n,{\cal Z})$ is generated by even numbers of $R_1$, $R_2$ and $R_3$. Thus, the possible set of group generators for $SL(n,{\cal Z})$ is 
\begin{align}
R_1^2, R_2^2,  R_3^2, \quad R_1 R_2, R_2 R_3, R_3 R_1, \quad R_2R_1, R_3R_2, R_1 R_3
\label{fullgenerator}\end{align}
with the properties that $R_2^2 = 1$ and $R_3^2 = 1$. A smaller set of the $SL(4,{\cal Z})$ generators is
\begin{align}
 R_1^2, R_1 R_3, R_2 R_3,
\end{align}
since other generators in (\ref{fullgenerator}) can be expressed with the generators in (\ref{mini}) through the following relations
\begin{align}
R_1 R_2 &= R_1 R_3 (R_2 R_3)^{-1}, \qquad R_2 R_1 = (R_1 R_2)^{-1} R_1^2\cr
R_3 R_2 &= (R_2 R_3)^{-1}, \qquad R_3 R_1 = (R_1 R_3)^{-1} R_1^2. \end{align}
Notice that 
\begin{align}
\{ R_1^2, R_1 R_3, R_2 R_3 \} = \{ U_1^2, U_1 U_3, U^{-1}_2\}.
\label{mini}\end{align}
{\it These three matrices generate $SL(4,{\cal Z})$.}
They can be shown to generate Trott's twelve generators $B_{ij}$\cite{Trott}.

Since we have tested the invariance of the zero mode partition function 
under $U_2$, we only need to check invariance under $U_1 U_3$ and $U_1^2$. 
For $U_1 U_3$, as previously we separate $U_1$ into
$U'$ and $M_3$,
\begin{align}
U_1 U_3 = U' M_3 U_3 = U' (M_3 U_3).
\end{align}
Since both $M_3$ and $U_3$ take ${Z}^{4d}_{\rm zero\, modes}$ to its complex conjugate,
$M_3 U_3$ is an invariance of the zero mode partition function. Thus from
(\ref{noma}),
\begin{align}
U_1 U_3:\hskip40pt Z_{zero\; modes}^{4d} \rightarrow |\widetilde \tau|^2  
Z_{zero\; modes}^{4d}. 
\end{align}
\vskip6pt

{\it $ {U_1}^2$ acts on ${Z}^{4d}_{\rm zero\; modes}$}

Since we have shown before 
\begin{align}
U_1:\hskip40pt Z_{zero\; modes}^{4d} \rightarrow |\widetilde \tau|^2  
Z_{zero\; modes}^{4d\; *},
\end{align}
then 
\begin{align}
{U_1}^2:\hskip40pt Z_{zero\; modes}^{4d} \rightarrow Z_{zero\; modes}^{4d}.
\end{align}
To summarize, we have
\begin{align}
U_2&:  Z_{\rm zero\; modes}^{4d} \rightarrow Z_{\rm zero\; modes}^{4d}, \cr
U_1 U_3&:  Z_{\rm zero\; modes}^{4d} \rightarrow |\widetilde \tau|^2 
Z_{\rm zero\; modes}^{4d}, \cr
{U_1}^2&: Z_{\rm zero\; modes}^{4d} \rightarrow  Z_{\rm zero\; modes}^{4d}.
\label{transSL2zero}\end{align}
One can derive a similar transformation property for 
$Z^{6d}_{\rm zero\;modes}$ using (\ref{zmpfsep}),
\begin{align}
&U_2:  Z_{\rm zero\; modes}^{6d} \rightarrow Z_{\rm zero\; modes}^{6d}, \cr
&U_1 U_3:  Z_{\rm zero\; modes}^{6d} \rightarrow |\widetilde \tau|^3 
Z_{\rm zero\; modes}^{6d}, \cr
&{U_1}^2: Z_{\rm zero\; modes}^{6d} \rightarrow  Z_{\rm zero\; modes}^{6d},
\label{transSL2zero6d}\end{align}
which follows from transformations on the factor $\epsilon$,
given in (\ref{epsil}). By inspection $\epsilon$ is invariant
under $U_2$ and $M_3$, and transforms as
\begin{align}
&U' :  \epsilon\rightarrow |\widetilde \tau| \epsilon. 
\end{align}
This can be seen by Poisson resummation since $\epsilon$ can be written as
\begin{align}
\epsilon 
=&\sum_{n_a}  {\rm exp}\{
-\frac{\pi R_6 \sqrt{\tilde g} }{R_1 R_2}g^{ab}
n_an_b - {\pi R_6\sqrt{\bar g}\over R_3R_1R_2|\widetilde\tau|^2}
\widetilde\gamma^a\widetilde\gamma^b n_an_b\}
\sum_{m, n_3}{\rm exp} \{-\pi 
(N+x)\cdot A \cdot (N+x)\},\cr
=&\; |\widetilde\tau|^{-1}\,U'\,\epsilon,
\end{align} 
where  
\begin{align}H_{12\alpha} &= n_\alpha,
\qquad H_{\alpha\beta\delta}
= {\epsilon_{\alpha\beta\delta}\over \tilde g}\, m,
\qquad m,n_\alpha\in {\cal Z}^4,\cr
A&=\left(\begin{matrix} {R_6\sqrt{\bar g}\over R_3R_1R_2}&i\gamma^3\cr
i\gamma^3&{R_6R_1R_2\over R_3\sqrt{\bar g}}\end{matrix}\right),\qquad
\det A = |\widetilde\tau|^2,\qquad
N=\left(\begin{matrix}n_3\cr m\end{matrix}\right),\qquad
x=\left(\begin{matrix}
\kappa^an_a +{\gamma^3\widetilde\gamma^an_a\over |\widetilde\tau|^2}\cr
i{R_6\sqrt{\bar g}\, \widetilde\gamma^3n_a\over R_3R_1R_2|\widetilde\tau|^2},
\end{matrix}\right).
\nonumber\end{align}
\vskip20pt

{\it $U'$ acts on ${Z}^{4d}_{\rm osc}$}

To derive how $U'$ acts on ${Z}^{4d}_{\rm osc}$,
we first separate the product on
$\vec n = (n,n_a)\ne \vec 0$ into a product
on (all $n$, but $n_\alpha\ne\nobreak (0,0)$)
and on ($n\ne 0$, $n_a = (0, 0))$.
Then using the regularized vacuum energy
(\ref{still}) expressed as sum over zero and non-zero transverse momenta
$p_\perp= n_a$
in (\ref{hsplit}), we find that 
(\ref{wpf4})  becomes
\begin{align}
Z^{4d, Maxwell}&=  Z^{4d}_{\rm zero\;modes} \cdot
\Big( e^{\pi R_6\over 6 R_3} \prod_{n\ne 0} {1\over
1 - e^{-2\pi  {R_6\over R_3}|n| -2\pi i \gamma^3 n}}\Big )^2
\cr& \hskip10pt \cdot
\Big( \prod_{n_a\in {\cal Z}^2\ne (0,0)} e^{-2\pi R_6 <H>_{p_\perp}}
\; \; \prod_{n^3\in {\cal Z}} {1\over 1 - e^{-2\pi
R_6\sqrt{g^{\alpha\beta} n_\alpha n_\beta} - 2\pi i\gamma^\alpha  n_\alpha}} \Big)^2.\cr
\label{wpf4again}\end{align}
As in \cite{DN} we observe the middle expression above can be
written in terms of the Dedekind eta function
$\eta(\widetilde \tau) \equiv e^{\pi i\widetilde\tau\over 12}
\prod_{n\in {\cal Z}\ne 0} (1-e^{2\pi i\widetilde\tau}),$ with
$\widetilde \tau = \gamma^3 + i{R_6\over R_3},$
\begin{align}
\Big( e^{\pi R_6\over 6 R_3} \prod_{n\ne 0} {1\over
1 - e^{-2\pi  {R_6\over R_3}|n| -2\pi i \gamma^3 n}}\Big )^2
= (\eta(\widetilde \tau)\bar\eta(\bar{\widetilde \tau}))^{-2}.
\label{eta5}\end{align}
This transforms under $U'$ in (\ref{5dmod}) as
\begin{align}
(\eta(-{\widetilde \tau}^{-1})\bar\eta(-\bar{\widetilde \tau}^{-1}))^{-2}
 = |\widetilde\tau|^{-2} \;
 (\eta(\widetilde \tau)\bar\eta(\bar{\widetilde \tau}))^{-2},
\label{etatt}\end{align}
where $\eta(-{\widetilde\tau}^{-1}) = (i\widetilde\tau)^{1\over 2}  
\eta(\widetilde\tau).$
In this way the anomaly of the zero modes in (\ref{noma})
is canceled by (\ref{etatt}). Lastly we demonstrate the third expression in 
(\ref{wpf4again})
is invariant under $U'$,
\begin{align}
\Big( \prod_{n_a\in {\cal Z}^2\ne (0,0)} e^{-2\pi R_6 <H>_\perp}
\; \; \prod_{n_3\in {\cal Z}} {1\over 1 - e^{-2\pi
R_6\sqrt{g^{\alpha\beta} n_\alpha n_\beta} - 2\pi i\gamma^\alpha n_\alpha}} 
\Big)^2 = PI,
\label{5mpf}\end{align}
where $PI$ is the modular invariant $2d$ path integral
of two massive scalar bosons of mass $\sqrt{\widetilde g^{ab}
n_a n_b},$ coupled to a worldsheet gauge field, on a two-torus 
in directions 3,6. Following \cite{DN}, with more detail in 
(\ref{btwo}),
we extract from (\ref{opf})
\begin{align}
Z^{4d}_{\rm osc}&=
( e^{-\pi R_6\sum_{\vec n\in {\cal Z}^3} \sqrt{g^{\alpha\beta}n_\alpha n_\beta}}
\; \; \prod_{\vec n\in {\cal Z}^3\ne \vec 0} {1\over 1 - e^{-2\pi
R_6\sqrt{g^{\alpha\beta} n_\alpha n_\beta} - 2\pi i \gamma^\alpha n_\alpha}} 
\Big)^2
\end{align}
the $2d$ path integral of free massive bosons coupling to the gauge field, 
where $n_a$ is fixed and non-zero,
\begin{align}
(PI)^{1\over 2}  &\equiv
e^{-\pi R_6\sum_{n_3\in {\cal Z}} \sqrt{g^{\alpha\beta}n_\alpha n_\beta}}
\; \; \prod_{n_3\in {\cal Z}} {1\over 1 - e^{-2\pi
R_6\sqrt{g^{\alpha\beta} n_\alpha n_\beta} +  2\pi i\gamma^\alpha n_\alpha}}\cr
&= \prod_{s\in {\cal Z}} {e^{-{\beta' E\over 2}}\over 1 -
e^{-\beta' E +  2\pi  i(\gamma^3 s + \gamma^a n_a)}}
\qquad \hbox{where}\; s\equiv n_3,\quad E\equiv \sqrt{{g}^{\alpha\beta}n_\alpha n_\beta},\quad
\beta'\equiv 2\pi R_6\cr
&=  \prod_{s\in {\cal Z}} {1\over \sqrt{2}\; \sqrt{\cosh{\beta' E}
- \cos{2\pi(\gamma^3 s + \gamma^a n_a)}}}\qquad
\hbox{for}\; n_a\rightarrow - n_a\cr
&= e^{-{1\over 2}\sum_{s\in{\cal  Z}} \big(\ln {[ \cosh\beta' E -
\cos{2\pi(\gamma^3 s + \gamma^a n_a)}]} + \ln{2}\big)}
\equiv  e^{-{1\over 2}\sum_{s\in{\cal  Z}} \;\nu(E)},
\label{newpi}\end{align}
where
\begin{align}
\sum_{s\in {\cal Z}} \nu(E) &\equiv
\sum_{s\in {\cal Z}} \big(\ln {[ \cosh\beta' E -
\cos{2\pi(\gamma^3 s + \gamma^a n_a)}]} + \ln{2}\big)\cr
&=\sum_{s\in {\cal Z}}\sum_{r\in {\cal Z}}
\ln\, {[{4\pi^2\over {\beta'}^2} ( r + \gamma^3 s + \gamma^a n_a)^2
+ E^2]}\label{eexpr}.\end{align}
We can show directly that
(\ref{eexpr}) is invariant under $U'$,
since
\begin{align}
&E^2 = g^{\alpha\beta}n_\alpha n_\beta = g^{33} s^2 + 2 g^{3 a} s n_a + g^{a b}
n_a n_b = {1\over R_3^2} (s + \kappa^an_a )^2 + \widetilde
g^{a b} n_a n_b,\cr
& {4\pi^2\over {\beta'}^2} ( r + \gamma^3 s + \gamma^a n_a)^2
= {1\over R_6^2} ( r + \widetilde\gamma^a n_a + \gamma^3 (s +
\kappa^a  n_a))^2,
\end{align}
then
\begin{align}
&{4\pi^2\over {\beta'}^2} ( r + \gamma^3 s + \gamma^a n_a)^2
+ E^2\cr & =
{1\over R_6^2} ( s + \kappa^a n_a)^2 \, |\widetilde\tau|^2
+ {1\over R_6^2} ( r + \widetilde \gamma^a n_a)^2
+ {2\gamma^3\over R_6^2} (  r + \widetilde \gamma^a n_a)
( s + \kappa^a n_a) +  \widetilde g^{a b} n_a n_b.
\label{argln}\end{align}
So we see the transformation $U'$ given in (\ref{5dmod}) leaves
(\ref{argln}) invariant if $s\rightarrow r$ and $r\rightarrow -s$.
Therefore (\ref{eexpr}) is invariant under $U'$, so that
$(PI)^{1\over 2}$ given in  (\ref{newpi}) is invariant under $U'$.
\vskip20pt

{\it  $M_3$ acts on ${Z}^{4d}_{\rm osc}$}

$M_3$ leaves the $Z^{4d}_{osc}$ invariant as can be seen from (\ref{wpf4again}) by shifting the integer $n_\alpha$ as
\begin{align}
n_3 \rightarrow -n_4, \qquad n_a \rightarrow n_{a+1}.
\end{align}
So, under $U_1 = U' M_3$,
\begin{align}
 Z_{\rm osc}^{4d} \rightarrow |\widetilde \tau|^{-2} Z_{\rm osc}^{4d}.
\end{align}
$U_2$ is an invariance of the oscillator partition function by 
inspection.
\vskip20pt
{\it  $U_3$ acts on ${Z}^{4d}_{\rm osc}$}

$U_3$ leaves the $Z^{4d}_{osc}$ invariant as can be seen from (\ref{wpf4again}) by shifting the integers $n_\alpha$ as
\begin{align}
n_3 \rightarrow -n_3, \qquad n_a \rightarrow n_{a}.
\end{align}
Thus, the oscillator partition function transforms under the $SL(4,{\cal Z})$ 
generators $\{ U^2_1, U_1 U_3, U_2\}$ as
\begin{align}
&\hskip10pt U_2:  Z_{\rm osc}^{4d} \rightarrow Z_{\rm osc}^{4d}, \cr
&U_1 U_3: Z_{\rm osc}^{4d} \rightarrow |\widetilde \tau|^{-2} Z_{\rm osc}^{4d}, \cr
&\hskip8pt {U_1}^2: Z_{\rm osc}^{4d} \rightarrow Z_{\rm osc}^{4d}.  
\label{transSL2osc}\end{align}
So together with (\ref{transSL2zero})
we have established invariance under (\ref{mini}), and thus proved 
the partition function for the $4d$ Maxwell theory
on $T^4$, given alternatively by (\ref{wpf4}) or (\ref{wpf4again}), 
is invariant under $SL(4,{\cal Z})$, the mapping class group of $T^4$.

\vskip20pt
{\it  $U'$ acts on $Z_{osc}^{6d}$}

For the $6d$ chiral theory on $T^2\times T^4$, 
where $<H>^{6d} \equiv {1\over 2}
\sum_{\vec p\in {\cal Z}^5}\sqrt{G_5^{lm} p_lp_m}$
appears in (\ref{ZHplusP6d}),
the $SL(3,{\cal Z})$ invariant regularized vacuum energy \cite{DN} becomes, 
\begin{align}
<H>^{6d}&= -{1\over 2\pi^4}\sqrt{G_5}\sum_{\vec n\ne\vec 0}
{1\over (G_{lm}n^ln^m)^3}\cr
&= -32\pi^2 \sqrt{G_5}\sum_{\vec n\ne \vec 0} {1\over (2\pi)^6
\Big( g_{\alpha\beta} n^\alpha n^\beta + (R_1^2 + R_2^2\beta^2\beta^2)
(n^1)^2 - 2\beta^2R^2_2 n^1n^2 + R_2^2 (n^2)^2  \Big)^3}\cr
\end{align}
and can be decomposed similarly to (\ref{hsplit}),
\begin{align}
<H>^{6d} &= \sum_{p_\perp\in {\cal Z}^4} <H>^{6d}_{p_\perp}
\;= \; <H>^{6d}_{p_\perp=0} + \sum_{p_\perp\in {\cal Z}^4\ne 0}
<H>^{6d}_{p_\perp},
\label{h6split}\end{align}
where 
\begin{align}
<H>^{6d}_{p_\perp} = -32\pi^2\sqrt{G_5}{1\over (2\pi)^4}
\int d^4y_{\perp}
e^{-ip_{\perp}\cdot
y_{\perp}} \sum_{n^3\in{\cal Z}\ne 0} {1\over |2\pi n^3 + y_{\perp}|^6},
\label{new6stew}\end{align}
with denominator $ |2\pi n^3 + y_{\perp}|^2 =
G_{33} (2\pi n^3)^2 + 2(2\pi n^3)G_{3k}y_\perp^k+G_{kk'}y_\perp^k y_\perp^{k'},$
with $k=1,2,4,5,$\begin{align}
<H>^{6d}_{p_\perp=0}&= -{1\over 12 R_3},\cr
<H>^{6d}_{p_\perp\ne 0}& = 
|p_\perp |^2 R_3 \sum_{n=1}^\infty
{\rm cos}(p_a \kappa^a 2\pi n)
\big [K_2 (2\pi n R_3 |p_\perp |) - K_0 (2\pi n R_3 |p_\perp |)\big]\cr
&= -\pi^{-1} |p_{\perp}| R_3\sum_{n=1}^{\infty} \cos(p_a \kappa^a 2\pi n) \frac{K_1(2\pi n R_3|p_\perp|)}{n}, \label{BES6}\end{align}
with $p_\perp = (p_1, p_2, p_a) = n_\perp = (n_1, n_2, n_a) 
= (n_1,n_2,n_4,n_5) \in {\cal Z}^4$, \hfill\break
$|p_\perp| = \sqrt{{(n_1)^2\over R_1^2} + 2\frac{\beta^2}{R_1^2}  
+ (\frac{1}{R_2^2} + \frac{{\beta^2}^2}{R_1^2}) n_2^2  
+ \widetilde g^{a b} n_a n_b}.$ \bigskip

The $U'$ invariance of (\ref{f6doscpf}) follows when 
we separate the product on $\vec n\in {\cal Z}^5\ne
\vec 0$ into a product on ($n_3 \ne 0,\; n_\perp \equiv (n_1,n_2,n_4,n_5)
= (0,0,0,0)$), and on
(all $n_3$, but $n_\perp = (n_1,n_2,n_4,n_5)\ne (0,0,0,0)$). Then
\begin{align}
Z^{6d}_{\rm osc}&= 
\bigl (e^{\pi R_6\over 6 R_3}
\prod_{n_3\in {\cal Z} \ne 0} {1\over
1 - e^{2\pi i \,(\gamma^3 \,n_3 + {i{R_6\over R_3} |n_3|\,)}}}\bigr )^3\cr
&\hskip20pt \cdot
\bigl(\prod_{n_\perp\in {\cal Z}^4\ne (0,0,0,0)}
e^{-2\pi R_6 < H>^{6d}_{p_\perp}} \prod_{n_3\in {\cal Z}}
{1\over 1-e^{-2\pi R_6\sqrt{ {\widetilde n}^2
+ g^{\alpha\beta} n_\alpha n_\beta}
\; +i2\pi\gamma^\alpha n_\alpha}}\bigr)^3\cr
&= \bigl ( \eta(\widetilde \tau)
\;\bar\eta (\bar{\widetilde \tau})\bigr)^{-3}\cr
&\hskip1pt
\cdot\bigl(\prod_{(n_1,n_2,n_4,n_5)\in {\cal Z}^4\ne (0,0,0,0)}
e^{-2\pi R_6 < H>^{6d}_{p_\perp}}
\prod_{n_3\in {\cal Z}}{1\over 1-e^{-2\pi R_6\sqrt{ g^{\alpha\beta} 
n_\alpha n_\beta + {\widetilde n}^2} \; 
+i2\pi\gamma^\alpha n_\alpha}  } \bigr)^3,\cr
\label{a6dpf}\end{align}
where $\widetilde \tau = \gamma^3 + i{R_6\over R_3}$, and
${\widetilde n}^2 \equiv {n^2_1\over R_1^2} + 2\frac{\beta^2}{R_1^2} n_1n_2 
+ (\frac{1}{R_2^2}+\frac{{\beta^2}^2}{R_1^2}) n_2^2.$ \hskip3pt
Under $U'$,
\begin{align}
\eta(\widetilde \tau)
\;\bar\eta (\bar{\widetilde \tau})\rightarrow |\widetilde\tau|\;
\eta(\widetilde \tau)\;\bar\eta (\bar{\widetilde \tau}).
\end{align}
$U'$ leaves invariant 
the part of the $6d$ oscillator 
partition function (\ref{a6dpf}) at fixed $n_\perp\ne 0$, since
\begin{align}
e^{-2\pi R_6 < H>^{6d}_{n_\perp\ne 0}}
\prod_{n_3\in {\cal Z}}{1\over 1-e^{-2\pi R_6\sqrt{ g^{\alpha\beta} 
n_\alpha n_\beta + {n^2_1\over R_1^2} +  2\frac{\beta^2}{R_1^2}n_1n_2  
+ (\frac{1}{R_2^2} + \frac{{\beta^2}^2}{R_1^2}) n_2^2 } \; 
+i2\pi\gamma^\alpha n_\alpha}  }\cr
\label{MB}\end{align}
is the square root of
the partition function on $T^2$ (now in the directions 3,6)
of a massive complex scalar
with $m^2 \equiv G^{11} n_1^2 + G^{22} n_2^2 +2G^{12} n_1 n_2 + 
\widetilde g^{a b}n_a n_b$, $4\le a, b \le 5$,
that couples to a constant gauge field
$A^\mu \equiv i G^{\mu i}  n_i$ with $\mu,\nu ={3,6};\,i,j = 1,2,4,5$.
The metric on this $T^2$ is $h_{33} = R_3^2\, , h_{66} = R_6^2 +
(\gamma^3)^2 R_3^2\, , \, h_{36} = -\gamma^3 R_3^2$. Its inverse is
$h^{3 3}= {1\over R_3^2} + {(\gamma^3)^2\over R_6^2}$, $h^{66}={1\over R_6^2}$
and $h^{3 6}={\gamma^3\over R_6^2}$. 
The manifestly $SL(2,{\cal Z})$ invariant path integral is

\begin{align} {\rm P.I.}
&=\int d\phi \,d\bar\phi \,\,
e^{-\int_0^{2\pi} d\theta^3 \int_0^{2\pi} d\theta^6\,\,
h^{\mu\nu}(\partial_\mu + A_\mu )\bar\phi
(\partial_\nu - A_\nu) \phi + m^2 \bar\phi \phi}\cr
&=\int\hskip-5pt d\bar\phi\, d\phi
e^{-\int_0^{2\pi} d\theta^3 \int_0^{2\pi} d\theta^6
\bar\phi (-({1\over R_3^2} + {(\gamma^3)^2\over R_6^2})\partial_3^2
-({1\over R_6})^2\partial_6^2 -2 {\gamma^3\over {R_6^2}}
\partial_3\partial_6 + 2A^3\partial_3 + 2A^6\partial_6
+ G^{11}n_1n_1 + G^{22} n_2 n_2 + 2G^{12} n_1 n_2 + G^{a b} n_a n_b )\phi}\cr
& = \det \Bigl (
[- ( {\scriptstyle{1\over R_3^2}} + ({\scriptstyle {\gamma^3\over R_6}})^2 )
\partial_3^2 - ({\scriptstyle {1\over R_6}})^2\partial_6^2
- 2\gamma^3 ({\scriptstyle {1\over R_6}})^2\partial_3\partial_6
+  G^{11}n_1n_1 + G^{22} n_2 n_2\cr &\hskip20pt+ 2 G^{12} n_1 n_2 + G^{a b} n_a  n_b
+ 2 iG^{3 a} n_a \partial_3
+ 2iG^{6 a} n_a \partial_6 ]\,\Bigr )^{-1}\cr
&= e^{- {\rm{tr}} {\rm{ln}} \Bigl[
- ( {\scriptstyle{1\over R_3^2}} + ({\scriptstyle {\gamma^3\over R_6}})^2 )
\partial_3^2 - ({\scriptstyle {1\over R_6}})^2\partial_6^2
- 2\gamma^3 ({\scriptstyle {1\over R_6}})^2\partial_3 \partial_6
+  G^{11} n_1 n_1 + G^{2 2} n_2 n_2 + 2G^{1 2} n_1 n_2 + G^{a b} n_a n_b + 2 iG^{3 a} n_a
\partial_3 + 2iG^{6 a}n_a \partial_6 \, \Bigr ] }
\cr &=e^{-\sum_{s\in {\cal Z}} \sum_{r\in {\cal Z}} \Bigl [ {\rm{ln}}
({4\pi^2 \over {\beta'}^2}r^2 +
( {\scriptstyle{1\over R_3^2}} + ({\scriptstyle {\gamma^3\over R_6}})^2 ) s^2
+ 2\gamma^3 ({\scriptstyle {1\over R_6}})^2 r s
+ G^{11} n_1 n_1 + G^{22} n_2 n_2 +2 G^{1 2} n_1 n_2 + G^{a b} n_a n_b
+ 2 G^{3 a}n_a \,s + 2G^{6 a} n_a\, r )
\, \Bigr ]}\cr
&= e^{-\sum_{s\in{\cal Z}}\nu(E)},
\label{btwo}\end{align}
where from (\ref{6dinverse}), $G^{11}= {1\over R_1^2},\,
G^{22} = {1\over R^2_2} + \frac{{\beta^2}^2}{R_1^2},\,
G^{1 2} = \frac{\beta^2}{R_1^2},\,
G^{ab} = g^{ab} + \frac{\gamma^a \gamma^b}{{R_6}^2},\hfill\break
G^{3a} = g^{3a} + \frac{\gamma^3 \gamma^a}{R_6^2},\,  
G^{6a} = {\gamma^a\over R_6^2}, G^{63} = \frac{\gamma^3}{R_6^2}$, and 
$\partial_3\phi = -i s \phi,\;
\partial_6\phi = -ir\phi$, $s=n_3$,
and $\beta'= 2\pi R_6$. 
The sum on $r$ is
\begin{align}\nu (E) &= 
 \sum_{r\in {\cal Z}} \ln \Bigl [{4\pi^2 \over {\beta'}^2}
(r + \gamma^3 s + \gamma^a n_a)^2
+ E^2 \Bigr ]\,,\label{bthree}\end{align}
with
$E^2 \equiv G^{lm}_5 n_l n_m = G_5^{11} n_1n_1 + G_5^{2 2} n_2 n_2 
+ G_5^{2 1} n_2 n_1 + G_5^{a b}
n_a n_b + 2 G^{a 3}_5 n_a n_3 + G_5^{3 3} n_3 n_3,$ 
and $G_5^{11} = {1\over R_1^2},\, G_5^{1 2} = \frac{\beta^2}{R_1^2},\,
G_5^{22} = {1\over R^2} +\frac{\beta^2\beta^2}{R_1^2},\, 
G_5^{1\alpha}=G_5^{2\alpha} = 0,\,
G_5^{3a} =  g^{3a} = {\kappa^a\over R_3^2},\hfill\break
G_5^{33} = g^{33} = {1\over R_3^2},\,
G_5^{ab} = g^{ab} = \widetilde g^{a b}
+ {\kappa^a \kappa^b\over R_3^2}.$
We evaluate the divergent sum $\nu(E)$ on $r$ by
\begin{align}{\partial \nu (E)\over \partial E} & =
\sum_r {2E\over {4\pi^2\over {\beta'}^2 } (r + \gamma^3 s +
\gamma^a n_a )^2 + E^2} \cr
& = \partial_E\ln \Bigl [ \cosh{\beta' E} -
\cos{2\pi \bigl ( \gamma^3 s + \gamma^a n_a \bigr)} \Bigr ],
\label{bfour}\end{align}
using the sum $\sum_{n\in {\cal Z}}
{2y\over {(2\pi n + z)^2 +y^2}} = {\sinh{y}\over {\cosh y -\cos z}}$.
Then integrating (\ref{bfour}), we choose the integration constant
to maintain modular invariance of (\ref{btwo}), 
\begin{align}\nu (E) = \ln \bigl [ \cosh{\beta' E} -
\cos{2\pi \bigl (\gamma^3 s  + \gamma^a n_a\bigr)} \bigr ]
+ \ln 2 \label{bfive}.\end{align}
It follows for $s=n_3$ that (\ref{btwo}) gives 
\begin{align}({\rm P.I.})^{1\over 2} &=
\prod_{s\in {\cal Z}} {1\over
\sqrt 2\sqrt {\cosh{\beta' E} -
\cos{2\pi ( \gamma^3 s + \gamma^a n_a )}}}\cr
&= \prod_{s\in {\cal Z}} {e^{-{\beta' E\over 2}}\over
{1 - e^{-\beta' E + 2\pi i ( \gamma^3 s + \gamma^a n_a )}}}\cr
&= e^{-\pi R_6 {\sum_{s\in {\cal Z}} \sqrt{G_5^{lm} n_l n_m}}}
\prod_{s\in {\cal Z}}
{1\over
{1 - e^{-2\pi R_6 \sqrt{G_5^{lm} n_l n_m}
+ 2\pi i  \gamma^3 s
+ 2\pi i \gamma^a  n_a}}}\,\cr
&= e^{-2\pi R_6 <H>_{n\perp}}
\prod_{n_3\in {\cal Z}}
{1\over
{1 - e^{-2\pi R_6 \sqrt{G_5^{lm} n_l n_m}
+ 2\pi i\gamma^3 n_3
+ 2\pi i \gamma^a n_a}}}\,,
\label{bfinal}\end{align}
which is (\ref{MB}).
Its invariance under $U'$ follows from the $U'$ invariance of 
(\ref{eexpr}), which differs from (\ref{bthree}) 
only by an additional contribution of ${\widetilde n}^2$ to the mass $m^2$.

Hence (\ref{bfinal}) and thus (\ref{MB}) are invariant under $U'$.

Furthermore $Z^{6d}_{\rm osc}$ is invariant under $M_3,U_2,U_3$ by 
inspection.

Using the same approach for proving $SL(4, {\cal Z})$ symmetry of the $4d$ 
partition function, we have shown the $6d$  oscillator partition function for 
the chiral boson given by  (\ref{ZHplusP6d}), or equivalently (\ref{a6dpf}), 
transforms as
\begin{align}
&\hskip10pt U_2:  Z_{\rm osc}^{6d} \rightarrow Z_{\rm osc}^{6d}, \cr
&U_1 U_3:  Z_{\rm osc}^{6d} \rightarrow |\widetilde \tau|^{-3} 
Z_{\rm osc}^{6d},\cr
&\hskip8pt {U_1}^2: Z_{\rm osc}^{6d} \rightarrow Z_{\rm osc}^{6d}. 
\label{transSL2osc}\end{align}
Together with (\ref{transSL2zero6d}), the $6d$ partition function 
$Z^{6d, chiral} \equiv Z^{6d}_{\rm zero\; modes} \,Z^{6d}_{\rm osc}$ is 
$SL(4,{\cal Z})$ invariant.

\vfill\eject
\providecommand{\MR}{\relax\ifhmode\unskip\space\fi MR }
\providecommand{\MRhref}[2]
{
}
\providecommand{\href}[2]{#2}

\end{document}